\documentclass[onefignum,onetabnum]{siamonline171218}

\usepackage{lipsum}
\usepackage{amsfonts}
\usepackage{graphicx,multirow}
\usepackage{epstopdf}
\usepackage{algorithmic}
\usepackage{amsmath,amssymb,advdate,mathrsfs,xfrac,dsfont,bm,stmaryrd}
\usepackage{caption,tabularx,systeme,xcolor,float,gensymb,graphicx,fancyvrb}
\usepackage{lipsum,mwe}   
\usepackage{comment}

\ifpdf
  \DeclareGraphicsExtensions{.eps,.pdf,.png,.jpg}
\else
  \DeclareGraphicsExtensions{.eps}
\fi

\usepackage{enumitem}
\setlist[enumerate]{leftmargin=.5in}
\setlist[itemize]{leftmargin=.5in}

\newsiamremark{remark}{Remark}
\newsiamremark{hypothesis}{Hypothesis}
\crefname{hypothesis}{Hypothesis}{Hypotheses}
\newsiamthm{claim}{Claim}

\headers{Adaptive importance sampling based on fault tree analysis}{Guillaume Chennetier, Hassane Chraibi, Anne Dutfoy, Josselin Garnier}

\title{Adaptive importance sampling based on fault tree analysis for piecewise deterministic Markov process}

\author{Guillaume Chennetier\footnotemark[1] \footnotemark[2]
\and Hassane Chraibi\thanks{EDF Lab Paris-Saclay, Boulevard Gaspard Monge, 91120 Palaiseau, France.}
\and Anne Dutfoy\footnotemark[1]
\and Josselin Garnier\thanks{CMAP, École Polytechnique, Institut Polytechnique de Paris, 91120 Palaiseau, France.}
}
\usepackage{amsopn}

\makeatletter
\newcommand*{\addFileDependency}[1]{
  \typeout{(#1)}
  \@addtofilelist{#1}
  \IfFileExists{#1}{}{\typeout{No file #1.}}
}
\makeatother

\newcommand*{\myexternaldocument}[1]{%
    \externaldocument{#1}%
    \addFileDependency{#1.tex}%
    \addFileDependency{#1.aux}%
}

\newcommand{\Uopt}{U_{\text{opt}}}
\newcommand{\Utheta}{U_{\bm{\theta}}}
\newcommand{\qopt}{q_{\text{opt}}}
\newcommand{\qtheta}{q_{\bm{\theta}}}
\newcommand{\thetaopt}{\bm{\theta}_{\text{opt}}}
\newcommand{\lambdaopt}{\lambda_{\text{opt}}}
\newcommand{\lambdatheta}{\lambda_{\bm{\theta}}}
\newcommand{\lambdazero}{\lambda_{0}}
\newcommand{\Kopt}{K_{\text{opt}}}
\newcommand{\Ktheta}{K_{\bm{\theta}}}
\newcommand{\Kzero}{K_{0}}
\newcommand{\tT}{\bm{\theta}\in\Theta}
\newcommand{\comp}[2]{\texttt{#1}_{\texttt{#2}}}
\newcommand{\Ua}[1]{U_{\bm{\theta}}^{\text{#1}}}
\newcommand{\nZ}{n_{\mathcal{Z}}}

\definecolor{changes}{rgb}{0.,0.,0.}

\ifpdf
\hypersetup{
  pdftitle={Adaptive importance sampling based on fault tree analysis for piecewise deterministic Markov process},
  pdfauthor={Guillaume Chennetier, Hassane Chraibi, Anne Dutfoy, and Josselin Garnier}
}
\fi

\myexternaldocument{ex_supplement}

\begin{document}

\maketitle

\begin{abstract}
  Piecewise deterministic Markov processes (PDMPs) can be used to model complex dynamical industrial systems.  The counterpart of this modeling capability is their simulation cost, which makes reliability assessment untractable with standard Monte Carlo methods. A significant variance reduction can be obtained with an adaptive importance sampling (AIS) method based on a cross-entropy (CE) procedure. The success of this method relies on the selection of a good family of approximations of the committor function of the PDMP. In this paper original families are proposed. Their forms are based on reliability concepts related to fault tree analysis: minimal path sets and minimal cut sets. They  are well adapted to high-dimensional industrial systems.  The proposed method is discussed in detail and applied to academic systems and to a realistic system from the nuclear industry. 
\end{abstract}


\section{Introduction}
\label{sec:intro}

The reliability assessment of industrial systems combines two issues: finding an appropriate framework to model these systems and proposing a method allowing the simulation of rare events since the failure probabilities are generally very low \cite{bucklew2004introduction,rubinoRareEventSimulation2009}. We focus here on hybrid dynamical industrial systems, i.e. systems whose time-dependent state is described by both continuous and discrete variables. 
The continuous variables are typically physical variables (such as temperature, pressure, or a level of liquid) that follow deterministic physical laws, and the discrete variables are typically the status of the components of the system that may be affected by random events.
Such a system can be modeled by a piecewise deterministic Markov process (PDMP) \cite{davisPiecewiseDeterministicMarkovProcesses1984,desgeorgesFormalismSemanticsPyCATSHOO2021,arismendiPiecewiseDeterministicMarkov2021,desaportaNumericalMethodsSimulation2015}. The modeling and simulation of hybrid systems is still an active field and other formalisms similar to PDMPs exist such as stochastic hybrid systems (SHS) \cite{polaStochasticHybridModels2003} or more recently general stochastic hybrid systems which generalize and encompass both PDMPs and SHS \cite{bujorianuGeneralStochasticHybrid}. The current work aims to enhance the efficiency of the \texttt{PyCATSHOO} toolbox \cite{chraibiPyCATSHOONewPlatform2016}, an EDF-developed computer code based on the PDMP formalism.

During the last decade, several attempts to adapt traditional methods of rare event simulation to hybrid systems have been proposed \cite{chraibiOptimalImportanceProcess2019,chraibiOptimalPotentialFunctions,krystulSequentialMonteCarloSimulation2005,turatiAdvancedRESTARTMethod2016,blomInteractingParticleSystembased2018,zulianiRareEventVerification2012,abateARCHCOMP21CategoryReport}. However, it still appears challenging to significantly reduce the required sample size compared to a standard Monte-Carlo method in the case of high-dimensional systems. Preliminary work \cite{chraibiOptimalImportanceProcess2019} established the connection between the optimal instrumental distribution of an importance sampling method for PDMPs and the use of the committor function of the process as an importance function (that we abbreviate IF). An importance function (also called reaction coordinate or collective variable in computational physics and chemistry \cite{McGibbon_2017}) offers a one-dimensional representation of the dynamics of a high-dimensional system. It associates to a given configuration of the system a real value that can be interpreted as a distance to a specific set of configurations. The committor function is known as the optimal IF to use for modeling phenomena in transition phase theory \cite{metznerTransitionPathTheory2009} and for splitting algorithms in rare event simulation \cite{cerouAdaptiveMultilevelSplitting2019}. It is the probability of realizing the rare event knowing the current state of the process. When dealing with stochastic processes taking values in standard Euclidean spaces, the committor function is frequently approximated by mixing explicit calculus on stochastic differential equations and machine learning methods \cite{liComputingCommittorFunctions2019,lucenteMachineLearningCommittor2019,khooSolvingHighdimensionalCommittor2018}. In our case, we can reduce the committor estimation problem to a parametric problem by looking for the best approximation of the committor function among a family of IFs that depend only on the status of the system components, which forms a discrete albeit high-dimensional variable. The failure of a specific component in a specific system configuration has to be encouraged to a greater or lesser extent depending on how it interacts with the other components. 
These interactions can be described by fault tree analysis through the concepts of minimal path sets and minimal cut sets, that are well-known in the reliability community  \cite{ruijtersFaultTreeAnalysis2015}. 
These concepts are often used to construct quantitative measures of system reliability such as importance indices for the components or approximations of the probability of failure \cite{rauzy1993new,lee1985fault,vcepin2011assessment}.
Here we will use them to design efficient importance sampling strategies. The construction of importance functions proposed in \cite{turatiAdvancedRESTARTMethod2016} to apply a RESTART method to hybrid systems is very close to our philosophy (see also \cite{buddeAutomatedRareEventSimulation2020} for the same idea with a splitting method for dynamic fault trees).

Given a family of IFs that serve as approximations of the committor function, each IF can be associated with an instrumental distribution for the importance sampling strategy. The search for the best candidate within this family is sequential using an adaptive importance sampling (AIS) method with a cross-entropy (CE) procedure (see \cite{bugalloAdaptiveImportanceSampling2017} for a global perspective on AIS, and \cite{deboerTutorialCrossEntropyMethod2005} for a general introduction on CE). Each iteration of the method consists of a simulation phase according to the current distribution and an optimization phase to refine this distribution. \textcolor{changes}{Classical parametric families of importance distributions (typically Gaussian mixtures in the literature) require a large number of parameters to sufficiently approximate the optimal distribution when it is quite complex. However it is well known that in high dimensions, importance sampling becomes tricky \cite{rubinsteinHowDealCurse2009} because of the degeneracy of the likelihood ratios.
This is one of the main concerns in the field and many recent works try to answer it when the optimal parameters of the instrumental distribution are sought by cross-entropy minimization \cite{elmasriOptimalProjectionImprove2022,masriImprovementCrossentropyMethod2020,uribeCrossentropybasedImportanceSampling2020,wangCrossentropybasedAdaptiveImportance2016}.}
In the sake of efficiency, the AIS literature is also paying increasing attention to recycling schemes for updating the instrumental distribution and/or estimating the final quantity by reusing samples from past iterations \cite{marinConsistencyAdaptiveImportance2019,delyonAsymptoticOptimalityAdaptive2018,cornuetAdaptiveMultipleImportanceSampling2011}. 
The paper \cite{delyonAsymptoticOptimalityAdaptive2018} further proves that the AIS estimator with a standard recycling scheme verifies a central limit theorem under appropriate assumptions. We give easily verifiable and physically interpretable conditions on the process and on the family of IFs that validate these assumptions. \\

\paragraph{Contribution} \color{changes}The major contribution of this work is to propose families of IFs suitable for approximating the committor functions of high-dimensional hybrid (industrial) systems. These families of IFs are easy to implement and interpret as they are constructed based on the ``minimal paths" and ``minimal cuts" of the system, which are classical concepts in reliability analysis. The original parameterization of these IFs also makes them highly flexible, allowing for easy refinement of the committor function approximation even in high dimension. 
We also provide detailed guidelines for the implementation of the AIS method with recycling scheme and prove the consistency and asymptotic normality of the associated IS estimator. \\ 

\color{black}
\paragraph{Structure} The paper is organized as follows.
\Cref{sec:intro} presents the notations and the industrial application case. It reminds the reader with the fundamental relation that exists between the committor function \(\Uopt\) of a piecewise deterministic Markov process and the optimal distribution \(\qopt\) of an importance sampling method for this process. 
We propose in \cref{sec:ApproxCommittor} three parametric families of functions \(\Utheta\) allowing to approximate the committor function \(\Uopt\) and to construct instrumental densities \(\qtheta\) producing low variance importance sampling estimators. These \(\Utheta\) functions are built on reliability properties of the system and we will explain for this purpose the notions of minimal path sets and minimal cut sets. 
 Our complete algorithm is given in \cref{sec:Algo} with its asymptotic properties. We propose an adaptive cross-entropy importance sampling algorithm with recycling of past samples (both for probability estimation and for updating the sampling policy). We also prove the consistency and the asymptotic normality of the estimator produced by the algorithm. 
Our recommendations for the implementation of the method are described in detail in \cref{sec:Implement}.
The method is applied in \cref{sec:Exp} to series/parallels systems and to two different configurations of the spent fuel pool system. Our method gives a dramatic variance reduction when estimating the probability of system failure even in the most complex case.
We finally discuss the implications and possible refinements of this work in \cref{sec:End}.

\subsection{Modeling hybrid systems with piecewise deterministic Markov processes}
\label{subsec:PrezPDMP}

A system failure is declared when continuous physical variables (e.g. temperature, pressure or liquid level in a tank) exceed critical threshold values. This only happens when key combinations of components fail and when repairs do not come in time. Component failures and repairs are then seen as random one-time events while the evolution of continuous variables is dictated by deterministic differential equations derived from physical laws. 
The high reliability of these systems is explained on the one hand by their high level of redundancy: the system can be reconfigured using several identical components to ensure its operation while waiting for the repair or replacement of broken components. On the other hand, the average waiting time before the failure of a component is generally considerably larger than the average waiting time before its repair. Such behavior lends itself very well to modeling by PDMPs.

PDMPs are a class of stochastic processes introduced and described by Mark Davis in 1984 \cite{davisPiecewiseDeterministicMarkovProcesses1984}. Apart from reliability considerations, these processes have since been used in many different areas, in particular to model chemical and biological phenomena \cite{lasotaStatisticalDynamicsRecurrent1992, lemaireExactSimulationJump2018,rudnickiPiecewiseDeterministicMarkov2015}. A considerable work has also been done on the theoretical properties of these processes, in particular on their long-time behaviors \cite{costaStabilityErgodicityPiecewise2008,azaisPiecewiseDeterministicMarkov2014}. They allow for a sophistication of Markov Chain Monte Carlo methods for the simulation of a posteriori distributions \cite{bouchardBouncyParticleSampler2018,bierkensZigZagProcessSuperEfficient2019,vanettiPiecewiseDeterministicMarkovChain2017}. 

A PDMP describes the evolution of one or more quantities over time. These quantities follow a deterministic trajectory that can change (jump) at random or deterministic times. Moreover, a PDMP is Markovian: the future of its trajectory depends only on its current state and the time already elapsed, but not on its past states. Working with PDMPs therefore requires managing two difficulties: first, its behavior is neither solely deterministic nor solely stochastic. Second, it is a hybrid process because it is composed of a continuous variable called "position" and a discrete variable called "mode".

The state of a PDMP at time \(t\) is denoted \(Z_t = \left(X_t,M_t\right) \in E\) where \(X_t \in \mathbb{X} \subset\mathbb{R}^{d_X}\) for some \(d_X \geq 1\) is the position and \(M_t\in\mathbb{M}\) is the mode of the PDMP (with \(\mathbb{M}\) a finite or countable set and \(E = \mathbb{X} \times \mathbb{M}\)). We consider in this work PDMPs of fixed duration \(t_{\max}>0\). We denote \(\mathcal{Z} = \left(Z_t\right)_{t\in[0,t_{\max}]}\) a complete trajectory and \(\mathcal{E}\) is the set of all possible trajectories of duration \(t_{\max}\) on \(E\) (the explicit description of the set \(\mathcal{E}\) is not necessary for the following but is detailed in \cite[section 1.2.3]{galtierAcceleratedMonteCarloMethods}). \textcolor{changes}{We will suppose in the following that one of the coordinates of the "position" variable $X \in \mathbb{X}$ is the total elapsed time and for any state $
z \in E$ we denote by $\tau_z$ this elapsed time (in particular for any $t>0$ we have $\tau_{Z_t} = t$)}. 

Given the state space \(E\), the behavior of a PDMP is characterized by three elements:

\begin{enumerate}
\item The flow \(\Phi\) that gives the deterministic trajectory.
\item The jump intensity \(\lambda\) which determines the distribution of the random jump times.
\item The transition kernel \(\mathcal{K}\) which determines the distribution of the post-jump locations.
\end{enumerate} 
\vspace{\baselineskip}

\color{changes}
\paragraph{Flow function} If there is no jump between time \(s\) and time \(s+h\), the mode $M_{s+h}$ of the PDMP remains constant and equal to $M_s$ and the position $X_{s+h}$ of the PDMP evolves in a deterministic way from $X_s$ as $\phi_{M_s}(X_s,h)$ where $\phi_m$ is
the solution of the differential equation $\dfrac{d \phi_m}{dh} = \bm{g}(\phi_m,m)$, $\phi_m(x,h=0)=x$. Here $\bm{g} : \mathbb{R}^{d_X}\times \mathbb{M} \rightarrow \, \mathbb{R}^{d_X} $ is a Lipschitz function. 
The trajectory of the process $(X_{s+h},M_{s+h})$ is then of the form $\Phi_{(X_s,M_s)}(h)$ where the flow function \(\Phi\) is defined by:
\begin{equation}
\mbox{For any \(z=(x,m) \in E\), }
    \Phi_z  : h \in [0,+\infty) \longmapsto \left(\phi_m(x,h),m\right) \in E.
\end{equation} 

\color{black}

\paragraph{Deterministic jumps}
By denoting \(E_m := \left\{z'=(x',m') \in E \ | \ m'=m \right\}\) we can write \(E = \bigcup_{m\in\mathbb{M}} E_m\). The boundary of the state space \(E_m\) for \(m \in \mathbb{M}\) is denoted by \(\partial E_m\). When the position reaches the boundary of the state space following the flow \(\Phi\), the PDMP jumps. Starting from a state \(z = (x,m) \in E_m\) at time \(s\) and assuming that no random jump takes place, the boundary \(\partial E_m\) is reached at a deterministic time \(t_z^{\partial} \in [0,+\infty] \): 
\begin{equation}
t_z^{\partial} = \inf \{h>0: \Phi_z(h) \in \partial E_m\} ,
\end{equation}
with the convention $\inf \emptyset = +\infty$. \\

\paragraph{Intensity function} 
For each state \(z\in E\), there is a random waiting time \(T_z\) at the end of which the process jumps (if it is smaller than the deterministic jump time \(t_z^{\partial}\)). The jump intensity \(\lambda\) is a function that associates to each state \(z\in E\) a weight \(\lambda(z) \in (0,+\infty)\). The larger \(\lambda(z)\) is, the more likely it is that the PDMP jumps when it passes through state \(z\). The distribution of \(T_z\) depends on \(\lambda\) through the following formula:
\color{changes}
\begin{equation}
\label{eq:jumpTime}
    \mathbb{P}\left(T_z > h \ | \ Z_{\tau_z} = z\right) = \mathds{1}_{h<t_z^{\partial}}\exp\left(-\int_0^h \lambda\left(\Phi_z(h')\right) dh' \right) .
\end{equation}
\color{black}

\paragraph{Transition kernel} Given the jump time and the state \(z^-\) from which the process jumps, the arrival state of the process after the jump is randomly chosen according to a Markovian transition kernel \(\mathcal{K}(z^-,\cdot)\). It is assumed for any \(z^-\) that the transition kernel \(\mathcal{K}(z^-,\cdot)\) admits a probability density function \(K\left(z^-,\cdot\right)\) with respect to a reference measure \(\nu_{z^-}\) on \(\mathscr{B}(E)\) (where \(\mathscr{B}(\cdot)\) indicates the Borelians of a set). So for any \(B \in \mathscr{B}(E)\):
\begin{equation}
\label{eq:jumpKernel}
    \mathcal{K}(z^-,B) = \int_B K\left(z^-,z\right) d\nu_{z^-}(z) .
\end{equation}

\paragraph{Probability density function of a trajectory} \textcolor{changes}{A fixed-time trajectory of a PDMP possesses a probability density function with respect to a dominant measure denoted by $\zeta$ which is a mixture of Lebesgue and Dirac measures. The definition of this dominant measure and the density of the PDMP trajectory can be found in Definition 3.2 and Theorem 3.3 of \cite{chraibiOptimalImportanceProcess2019}, respectively. Using our notations, the expression for this density is as follows}. Let \(\mathcal{Z} = \left(Z_t\right)_{t\in[0,t_{\max}]}\) be a PDMP trajectory and \(\nZ\) its number of jumps. We denote by \(z_0\) the initial state of the trajectory, by \(t_0\) the waiting time before the first jump, and for \(k=1,\dots,\nZ-1\) we denote by \(z_k\) the state of the process after the \(k\)-th jump and by \(t_k\) the waiting time between the \(k\)-th and the \((k+1)\)-th jump. Finally let \(t_{\nZ} = t^{\partial}_{z_{\nZ}} = t_{\max} - \sum_{k=0}^{\nZ-1} t_k\). The probability density function of this trajectory can then be expressed as:
\begin{equation}
\label{eq:Density}
     \pi_{\lambda,K}(\mathcal{Z}) = \prod_{k=0}^{\nZ} \left[\lambda\left(\Phi_{z_k}(t_k) \right)\right]^{\mathds{1}_{t_k<t_{z_k}^{\partial}}} \times \exp\left[-\int_0^{t_k} \lambda\left(\Phi_{z_k}(u) \right)\, \text{d} u\right]  \times \prod_{k=0}^{\nZ-1} K\left(\Phi_{z_k}(t_k),z_{k+1}\right) .
\end{equation}
We consider the state space \(E\) and the flow \(\Phi\) of the PDMP fixed once and for all in this paper. A distribution \(\pi_{\lambda,K}\) of a PDMP trajectory on \(\mathcal{E}\) can thus be totally determined by the choice of the jump intensity \(\lambda\) and the jump kernel density \(K\).  \\

\color{changes}
\paragraph{System failure} In the case of industrial systems, the position of the PDMP contains the physical variables that determine the failure of the system and if necessary all the variables allowing the process to be Markovian such as the elapsed time. 
The physical variables evolve in time according to the flow \(\Phi\) given by physical laws. The mode of the PDMP contains the status of each component (active, inactive, broken, etc.). The failure and repair rates of each component (which can depend on the value of the physical variables) determine the jump intensity \(\lambda\) and the jump kernel density \(K\) (see \cref{sec:Exp} for a numerical example). \\

The system fails when the position reaches a critical region $\mathbb{X}_D$ determined by threshold values for the physical variables. The critical region $\mathbb{X}_D$ can only be reached in certain modes (the critical thresholds cannot be reached by the physical variables when all components are functioning, for example). Let 
$\mathbb{M}_D$ denote the set of modes $m$ such that the position can reach $\mathbb{X}_D$ by following the flow $\phi_m$ corresponding to the mode $m$. The set of states corresponding to a failed system is therefore $D = \mathbb{X}_D \times \mathbb{M}_D \subset E$. The set of faulty trajectories is written \(\mathcal{D}\subset \mathcal{E}\) and corresponds to:
\begin{equation}
\mathcal{D} := 
    \left\{\left(Z_t\right)_{t\in[0,t_{\max}]} \in \mathcal{E} \ : \ \exists \, t \in[0,t_{\max}] \text{ such that } Z_t \in D  \right\} .
\end{equation}

\paragraph{Objective} 
We denote by \(\lambdazero\), \(\mathcal{K}_{0}\) and \(\Kzero\) the jump intensity, jump kernel and jump kernel density of the PDMP modeling the system whose failure probability we wish to estimate, and \(\pi_{0} \equiv \pi_{\lambdazero,\Kzero}\) the corresponding PDMP distribution. Our objective is to estimate the following probability:
\begin{equation}
    P := \mathbb{P}_{\pi_{0}}\left(\mathcal{Z} \in \mathcal{D} \right) = \mathbb{E}_{\pi_{0}}\left[\mathds{1}_{\mathcal{Z}\in\mathcal{D}} \right] .
\end{equation}

\color{black}
\paragraph{Simulation cost} Most of the computational cost for complex industrial systems comes from solving the differential equations that define the flow of the PDMP. We consider in the following that simulating several tens of uniform random variables and evaluating the density of a PDMP trajectory have a negligible cost compared to the computation of the flow between two jumps. This will guide the choice of our jump time simulation method and our optimization strategy for AIS (both are described in \cref{par:Simu}).
Note that, even if the flow appears in the formula of the density of a trajectory \cref{eq:Density}, it does not have to be computed again since it was already evaluated to generate the trajectory. \\

\begin{figure}[h!]
\begin{center}
    \includegraphics[scale=0.15]{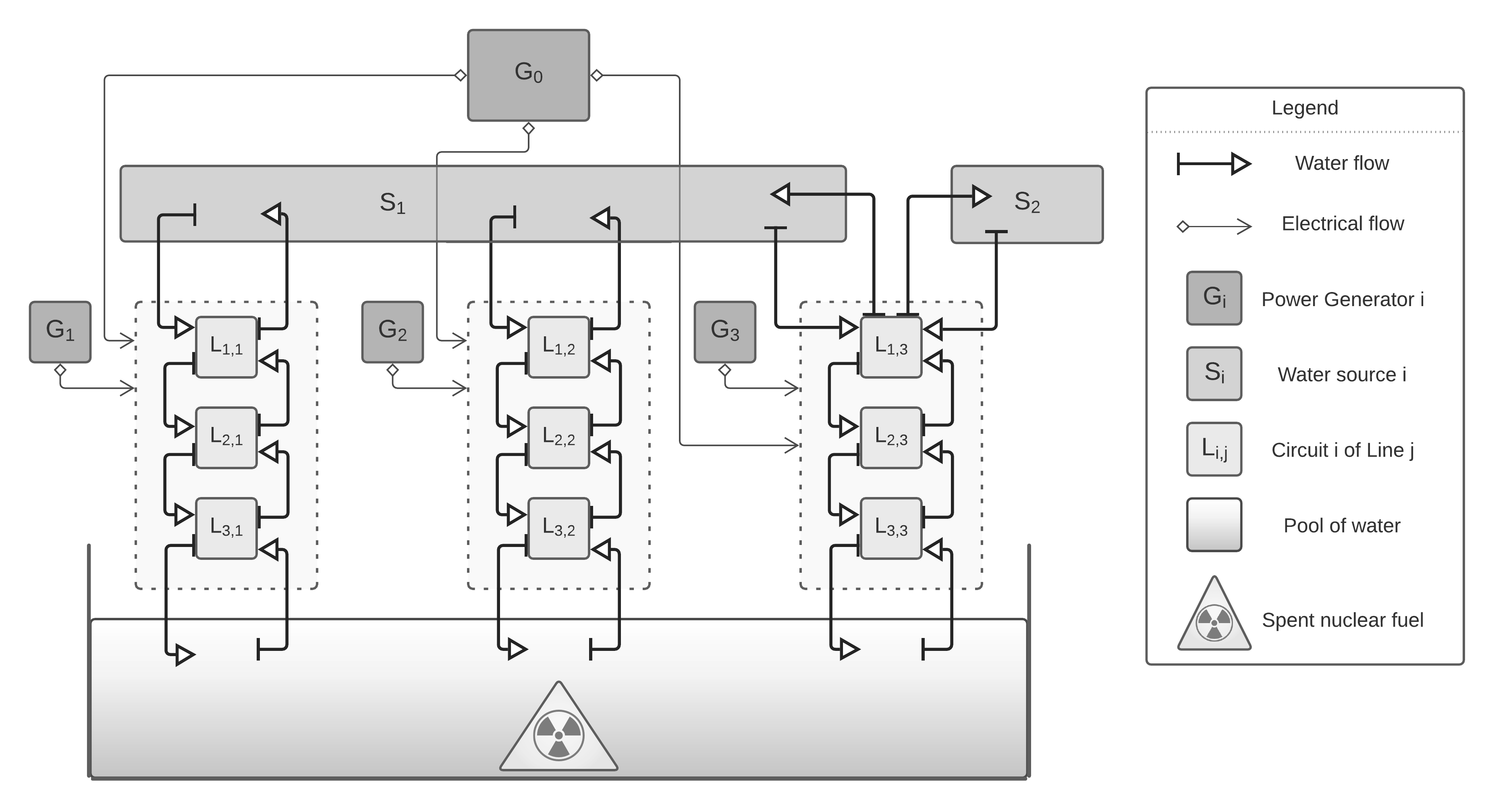}
    \caption{\small\textbf{Representation of the spent fuel pool.} The temperature of an external cold water source \(\texttt{S}_1\) is transfered to the pool by means of three sealed circuits connected by heat exchangers \(\texttt{L}_{1,1}\), \(\texttt{L}_{2,1}\) and \(\texttt{L}_{3,1}\) forming a line \(\texttt{L}_1\). The system has a general power supply \(\texttt{G}_0\). In the event of a problem with one of these components, the system is equipped with two other lines \(\texttt{L}_2\) and \(\texttt{L}_3\) identical to \(\texttt{L}_1\), an emergency diesel generator for each line \(\texttt{G}_1\), \(\texttt{G}_2\) and \(\texttt{G}_3\), and a second outside water source \(\texttt{S}_2\) accessible only to the third line \(\texttt{L}_3\).}
    \label{fig:SFP_Representation}
\end{center}
\end{figure}

\paragraph{Test case: the spent fuel pool} We propose a redesigned version of the system presented in \cite{chraibiPyCATSHOONewPlatform2016}. It is a simplification of the real operation of the storage pools of water for spent fuel from nuclear reactors. The water in the pool cools the fuel and provides protection from radiation. Conversely, the fuel heats the water in the pool, which will eventually boil, evaporate and allow the fuel to damage the structure and contaminate the outside environment. System failure is declared when the water level in the pool has reached a critical level. All the components of the system (shown in \cref{fig:SFP_Representation}) are designed to keep the water in the pool cold enough to prevent it from evaporating. 

On this system the position of the PDMP corresponds to the water temperature and the water level in the pool; its mode is the combination of the status  (active, inactive, broken) of the components. The domain \(D\) is defined by all the states of \(E\) in which the water temperature (first coordinate of the position) is \(100^{\circ}\text{C}\) and the water level (second coordinate of the position) is lower or equal to the critical threshold. We notice that these positions are not accessible for the flow for all modes, for example if no component is broken. If on the other hand, all the generators \(\texttt{G}_0,\texttt{G}_1,\texttt{G}_2\) and \(\texttt{G}_3\) are and remain broken for example, then the flow reaches one of these positions in finite time. A more formal description of the modes allowing the flow to bring the process into \(D\) is given in \cref{sec:ApproxCommittor}. This description will turn out to be important to build a good estimator of the failure probability. The system data allowing to compute the flow, and the different failure and repair rates of each component allowing to build the jump intensities and kernels are given in \cref{sec:Exp}.  

\subsection{Importance sampling with piecewise deterministic Markov processes}

\label{subsec:PrezIS}
We are interested in estimating \(P\) the probability that the system fails before \(t_{\max}\). In the nuclear and hydraulic energy sector, the reliability requirements for the systems are very high. The system failure is therefore a rare event with a probability of \(10^{-5}\) or smaller. \\

\paragraph{Crude Monte-Carlo estimator (CMC)} It is well known that CMC methods are quite inefficient in this case because an overwhelming majority of realizations will not produce any failure. A CMC estimator of \(P\) from an \textit{i.i.d} sample \(\mathcal{Z}_1,\dots,\mathcal{Z}_N \sim \pi_{0}\) is given by:
\begin{equation}
    \widehat{P}^{\text{CMC}}_N  := \dfrac{1}{N} \sum_{k=1}^N \mathds{1}_{\mathcal{Z}_k\in\mathcal{D}} \xrightarrow[N\rightarrow\infty]{\text{a.s.}}  \mathbb{E}_{\pi_{0}}\left[\mathds{1}_{\mathcal{Z} \in \mathcal{D}} \right] = P \, .
\end{equation}

An accurate CMC estimator of \(P\) (say with a coefficient of variation of 0.1) requires about \(\sfrac{100}{P}\) realizations. Each realization implies to simulate a PDMP trajectory which is very expensive especially for large industrial systems (as previously mentioned, the deterministic parts of the trajectory correspond to complex physical phenomena and result from the resolution of expensive computer codes). It is therefore unthinkable to simulate several tens of millions of trajectories to estimate the probability of system failure. \\

\paragraph{Importance sampling estimator (IS)} Introduced in 1951 by Kahn and Harris \cite{kahn1951estimation}, importance sampling is used to estimate the expectation of a random quantity under an arbitrary distribution. In the context of rare event simulation, this method allows to estimate the probability \(P\) of the rare event using an instrumental distribution \(q\) whose support is included in the support of the original distribution \(\pi_{0}\) and that realizes the event more frequently than it. See \cite{bucklew2004introduction} for a rare event perspective of IS, \cite{tabandeh2022review} for a recent review of IS methods in reliability assessment and \cite{elviraAdvancesImportanceSampling2021} for recent advances in IS for more general purpose.
From the formula
\begin{align}
    P &= \mathbb{E}_{\pi_{0}}\left[\mathds{1}_{\mathcal{Z}\in\mathcal{D}} \right] = \int_{\mathcal{E}} \mathds{1}_{\mathcal{Z}\in\mathcal{D}} \pi_{0}(\mathcal{Z})\,d\zeta(\mathcal{Z}) \nonumber \\
    &= \int_{\mathcal{E}} \mathds{1}_{\mathcal{Z}\in\mathcal{D}} \dfrac{\pi_{0}(\mathcal{Z})}{q(\mathcal{Z})}q(\mathcal{Z})\,d\zeta(\mathcal{Z}) = \mathbb{E}_{q}\left[\mathds{1}_{\mathcal{Z}\in\mathcal{D}}\dfrac{\pi_{0}(\mathcal{Z})}{q(\mathcal{Z})} \right] ,
\end{align}
we can deduce the form of the importance sampling estimator of \(P\) with an \textit{i.i.d} sample \(\mathcal{Z}_1,\dots,\mathcal{Z}_N \sim q\):
\begin{equation}
\label{eq:estimIS}
    \widehat{P}_N^{\text{IS}} := \dfrac{1}{N} \sum_{k=1}^N \mathds{1}_{\mathcal{Z}_k\in\mathcal{D}} \dfrac{\pi_{0}(\mathcal{Z}_k)}{q(\mathcal{Z}_k)} \xrightarrow[N\rightarrow\infty]{\text{a.s.}} \mathbb{E}_{q}\left[\mathds{1}_{\mathcal{Z} \in \mathcal{D}} \dfrac{\pi_{0}(\mathcal{Z})}{q(\mathcal{Z})}\right] = P \, .
\end{equation}
\paragraph{Optimal importance process for PDMPs}
\label{OptIS}
The variance of \(\widehat{P}_N^{\text{IS}}\) strongly depends on the choice of \(q\). The optimal distribution \(\qopt : \mathcal{Z} \mapsto \frac{1}{P} \mathds{1}_{\mathcal{Z}\in\mathcal{D}} \,\pi_{0}(\mathcal{Z})\) produces a zero variance IS estimator. Although inaccessible in practice, this form guides us on the choice of the instrumental density to use. We also know from \cite{chraibiOptimalImportanceProcess2019} that the process of distribution \(\qopt\) is a PDMP  with the same state space and with the same deterministic flow as the original PDMP of distribution \(\pi_{0}\).

The optimal jump intensity \(\lambdaopt\) and optimal jump kernel density \(\Kopt\) given in \cref{eq:OptIntKer} can be expressed in terms of the committor function of the process \(\Uopt\). It is defined here as follows for any states \(z^-,z \in E\) by:
%
%
\color{changes}
\begin{equation}
\label{eq:Uopt}
    \Uopt(z) := \mathbb{E}_{\pi_{0}}\left[\mathds{1}_{\mathcal{Z}\in\mathcal{D}}\mid Z_{\tau_z} =z\right] \quad \text{and} \quad \Uopt^- (z^-) := \int_E \Uopt(z) \,\Kzero\left(z^-,z\right) d\nu_{z^-}(z) .
\end{equation}
\color{black}
\(\Uopt\) represents the probability of reaching \(\mathcal{D}\) knowing the current state of the trajectory and \(\Uopt^-\) represents the same quantity with the additional assumption that the process jumps immediately. Theorems 4.3 and 4.4 of \cite{chraibiOptimalImportanceProcess2019} give us the following result:

\color{changes}
\begin{theorem}[Optimal jump intensity and jump kernel]
 For states \(z^-,z\in E\), the optimal jump intensity and optimal jump kernel density are given by:
\begin{equation}
\label{eq:OptIntKer}
 \lambdaopt(z) = \lambdazero(z) \times \dfrac{\Uopt^- \left(z\right)}{\Uopt \left(z\right)}  \quad \text{and} \quad  \Kopt \left(z^-,z\right) = \Kzero \left(z^-,z\right) \times \dfrac{\Uopt \left(z\right)}{\Uopt^- \left(z^-\right)}  .
\end{equation}   
We have $q_{\lambdaopt,\Kopt}(\mathcal{Z}) = \qopt(\mathcal{Z}) = \frac{1}{P} \mathds{1}_{\mathcal{Z}\in\mathcal{D}} \,\pi_{0}(\mathcal{Z})$ and this distribution produces a zero-variance IS estimator of $P$.
\end{theorem}

%
\color{black}
\textbf{Remark:} understanding these equations gives a good intuition of the behavior of the optimal process. If from a given state \(z^-\in E\): 
\begin{enumerate}
    \item the process is \(k\) times more likely to reach \(D\) before the end of the simulation by jumping now than by not jumping, then the optimal intensity \(\lambdaopt\) must be \(k\) times larger than the nominal intensity \(\lambdazero\),
    \item and if the process is \(k\) times more likely to reach \(D\) before the end of the simulation by jumping to state \(z\) than by jumping randomly according to \(\mathcal{K}\), then the value of the optimal jump kernel density \(\Kopt\) between \(z^-\) and \(z\) must be \(k\) times larger than that of the nominal jump kernel density \(\Kzero\).
\end{enumerate}
The optimal distribution \(\qopt\) is thus completely characterized by the original distribution \(\pi_{0}\) and by the committor function \(\Uopt\).

\section{Parametric approximation of the committor function \(\Uopt\)}
\label{sec:ApproxCommittor}

In practice the true committor function is not accessible. The best we can do is to replace it by a function that we call importance function (IF) and whose behavior is as close as possible to \(\Uopt\). Rather than trying to learn the committor function among a nonparametric  class of functions, we look for the best approximation of \(\Uopt\) among a parametric family of IFs \(\big(\Utheta\big)_{\tT}\). Here \(\tT \subset \mathbb{R}^{d_{\bm{\theta}}}\) is a parameter of dimension \(d_{\bm{\theta}}\). Each IF \(\Utheta\) is associated with a PDMP importance distribution \(\qtheta \equiv \pi_{\lambdatheta,\Ktheta}\) whose jump intensity \(\lambdatheta\) and jump kernel \(\Ktheta\) are defined by:
\color{changes}
\begin{equation}
    \label{eq:thetaIntKer}
 \lambdatheta(z) = \lambdazero(z) \times \dfrac{\Utheta^-\left(z\right)}{\Utheta\left(z\right)} \quad \text{and} \quad \Ktheta\left(z^-,z \right) = \Kzero\left(z^-,z\right) \times \dfrac{\Utheta\left(z\right)}{\Utheta^-\left(z^-\right)}     .
\end{equation}
%

\color{black}
The more faithful is the approximation \(\Utheta\) to the committor function \(\Uopt\), the closer the instrumental distribution \(\qtheta\) should be to the optimal distribution \(\qopt\), and the larger the variance reduction of the IS method should be. The committor function can be interpreted as a proximity measure between a state of the process and the set \(D\). \\ 
%

\paragraph{IFs without position dependency} Since it is sufficient to stay long enough in \(\mathbb{M}_{D}\) to end up in \(D\), the main obstacle to overcome in order to realize the rare event is to reach \(\mathbb{M}_{D}\). Moreover, the committor function appears only as ratios evaluated at identical positions but distinct modes (see \cref{eq:thetaIntKer}), which removes some of the position dependence. It is therefore reasonable to restrict ourselves to IFs which depend only on the mode and not on the position 
of the process.
It remains to determine how to quantify the proximity of a mode (which represents the status of the system components) to the set \(\mathbb{M}_{D}\). For this purpose we will exploit a static and Boolean representation of the system, and make use of concepts from the reliability theory: the minimal path sets (MPS) and minimal cut sets (MCS).

\subsection{Minimal path sets (MPS) and Minimal cut sets (MCS)}

\color{changes}
\label{subsec:MPSMCS}
In the static point of view, we consider the final mode of the trajectory and we declare that the trajectory has failed if that mode belongs to \(\mathbb{M}_{D}\) (without taking account the time during which the position evolves to \(D\) when its mode belongs to \(\mathbb{M}_{D}\) as well as the possibility that the system is repaired during this time).
The path sets and cut sets can be defined as follows.
\begin{itemize}
    \item The path sets are the sets of components whose operation prevents the system failure.
    \item The cut sets are the sets of components whose malfunction causes the system failure.
\end{itemize} 
A path/cut set is minimal if it contains no other path/cut set. \textcolor{changes}{These concepts can be understood very well with examples}. \\ 

\begin{figure}[!htb]
   \begin{minipage}{0.47\textwidth}
     \centering
     \includegraphics[width=.8\linewidth]{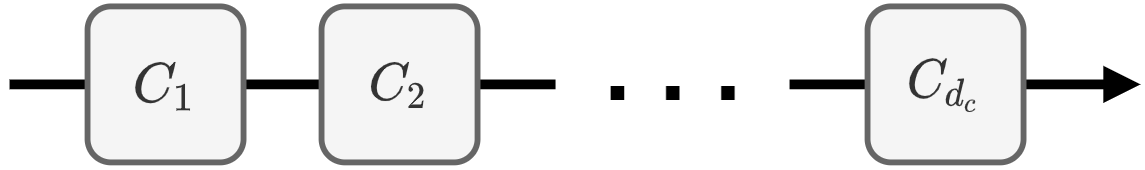}
     \caption{\small\textbf{Series system} with $d_{\bm{c}}$ components. It has one MPS~: $\{ {\texttt C}_{1}, \ldots , {\texttt C}_{{d_{\bm{c}}}} \}$ and $d_{\bm{c}}$ MCS~: $ \{ {\texttt C}_{1} \}, \ldots, \{ {\texttt C}_{d_{\bm{c}}} \}$.}
     \label{fig:seriesSystem}
   \end{minipage}\hfill
   \begin{minipage}{0.49\textwidth}
     \centering
     \includegraphics[width=.6\linewidth]{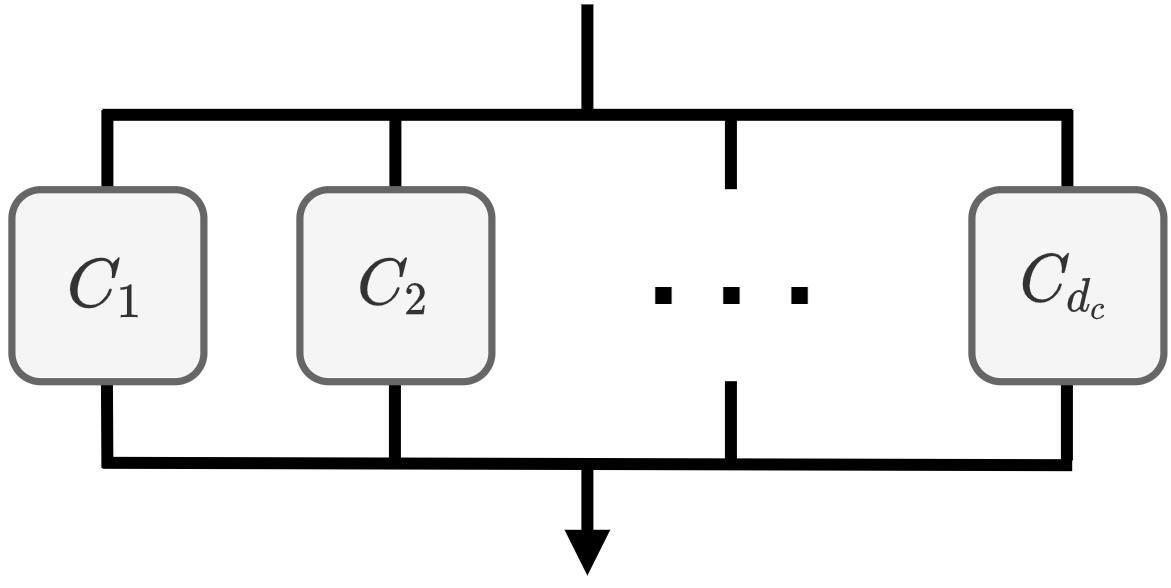}
     \caption{\small\textbf{Parallel system} with $d_{\bm{c}}$ components. It has $d_{\bm{c}}$ MPS~: $\{ {\texttt C}_{1} \}, \ldots,\{ {\texttt C}_{d_{\bm{c}}}\}$ and one MCS~: $\{ {\texttt C}_{1},\ldots, {\texttt C}_{d_{\bm{c}}}\}$.}
     \label{fig:parallelSystem}
   \end{minipage}
\end{figure}

\paragraph{Series and parallel systems} Here are two extreme examples of industrial systems to keep in mind for the following. A series system is a configuration of components in which the failure of any one component is sufficient to cause system failure (see Figure~\ref{fig:seriesSystem}). A parallel system is a configuration of components in which the failure of all components is necessary to cause system failure (see Figure~\ref{fig:parallelSystem}). 
A series system has only one MPS and a parallel system has \(d_{\bm{c}}\) MPS. In contrast, a series system has \(d_{\bm{c}}\) MCS and a parallel system has only one MCS.\\

\paragraph{Spent fuel pool example}
The test case described in \cref{subsec:PrezPDMP} can be represented as a series/parallel diagram (see Figure~\ref{fig:poolPS}) facilitating its decomposition into MPS and MCS. The MPS correspond to all vertical combinations and the MCS to all horizontal combinations.\\

\begin{center}
    \includegraphics[scale=0.4]{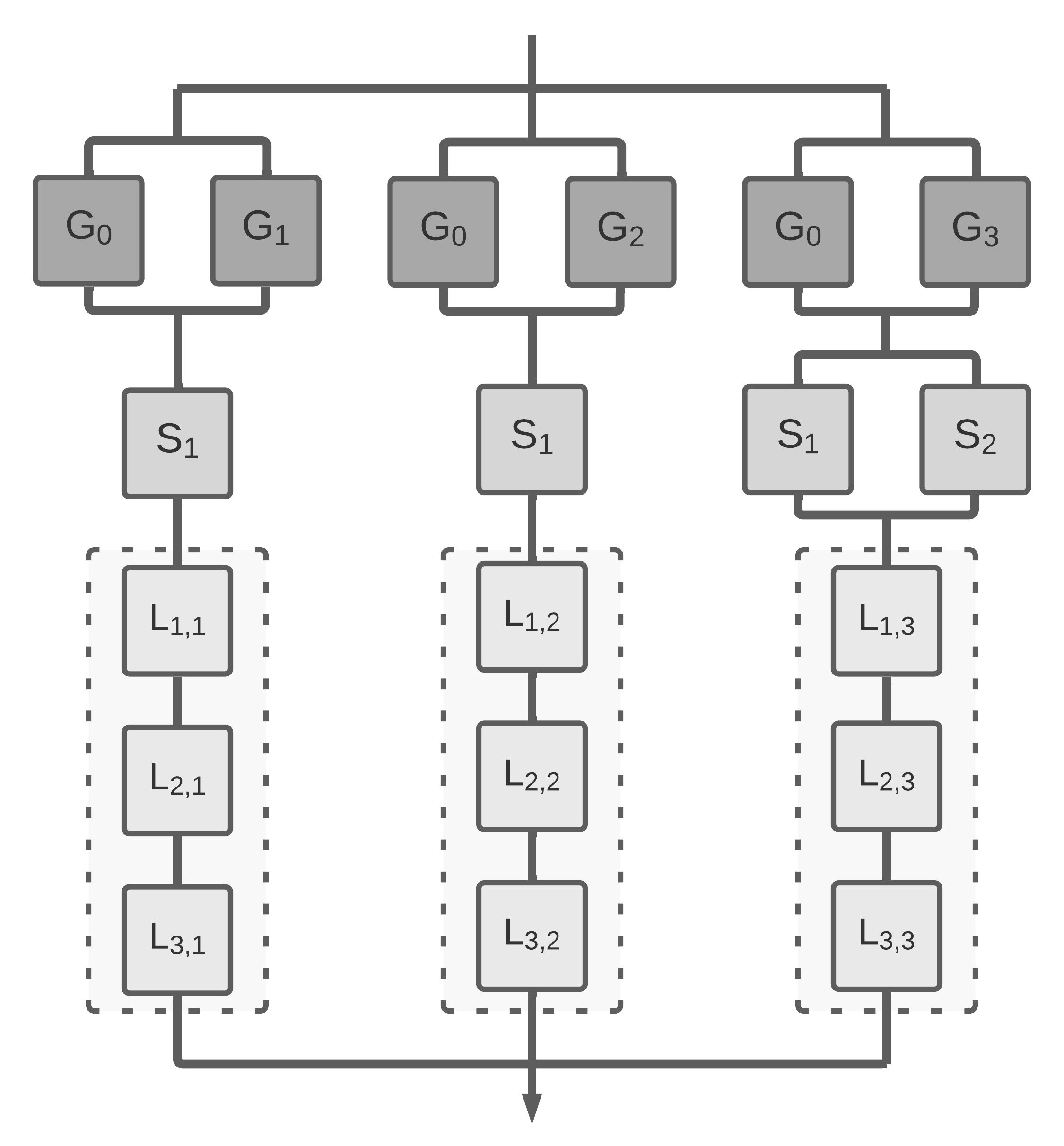}
    \captionsetup{singlelinecheck=off}
    \captionof{figure}{\small\textbf{Series/Parallel diagram of the spent fuel pool of Figure~\ref{fig:SFP_Representation}.} There are 8 MPS and 69 MCS in this system.
 MPS examples: \( \{ \comp{G}{0},\comp{S}{1},\comp{L}{1,1},\comp{L}{2,1},\comp{L}{3,1} \} \) and \( \{ \comp{G}{0},\comp{S}{2},\comp{L}{1,3},\comp{L}{2,3},\comp{L}{3,3} \}\).
 MCS examples: \( \{ \comp{L}{1,1},\comp{L}{3,2},\comp{L}{1,3} \} \) and \( \{ \comp{G}{0},\comp{S}{1},\comp{G}{3} \} \).
   }
   \label{fig:poolPS}
\end{center}

\paragraph{Formal definition with Boolean algebra}


Let \(d_{\bm{c}}\in\mathbb{N}^*\) and \(\bm{c} = (c_1,\dots,c_{d_{\bm{c}}}) \in \{0,1\}^{d_{\bm{c}}}\), we note \(\llbracket 1,d_{\bm{c}}\rrbracket = \{1,\dots,d_{\bm{c}}\}\) and for any \(I \subset \llbracket 1,d_{\bm{c}}\rrbracket\): 
\[
\bigvee_{j\in I} c_j = \max \left\{ c_j, \, j\in I \right\} \quad \text{and} \quad \bigwedge_{j\in I} c_j = \min \left\{c_j, \, j\in I \right\} .
\] 

Let \(\bm{c}^{\prime} = (c_1^{\prime},\dots,c_{d_{\bm{c}}}^{\prime}) \in\{0,1\}^{d_{\bm{c}}}\), we write \(\bm{c}\leq\bm{c}^{\prime}\) if and only if \(c_j\leq c_j^{\prime}\) for all \(j\in \llbracket 1,d_{\bm{c}}\rrbracket\). Let \(\varphi\) be a Boolean function, i.e from \(\{0,1\}^{d_{\bm{c}}}\) to \(\{0,1\}\). The function \(\varphi\) is non-decreasing if and only if \(\varphi(\bm{c})\leq\varphi(\bm{c}^{\prime}) \Longleftrightarrow \bm{c}\leq\bm{c}^{\prime}\) for all \(\bm{c},\bm{c}^{\prime} \in \{0,1\}^{d_{\bm{c}}}\). 
Combining Theorem 1.14 and Theorem 1.21 from \cite{Crama2011BooleanF}, we obtain the following decomposition:
\begin{theorem}[Complete disjonctive normal form of an non-decreasing Boolean function]
\label{th:DNF}
If \(\varphi\) is an non-decreasing Boolean function, then it admits a unique decomposition (except for the numbering of the terms) of the form: 
\begin{equation}
    \varphi(\bm{c}) = \bigvee_{i=1}^{d_{\varphi}} \bigwedge_{j\in I_i} c_j \quad \text{for all } \bm{c}\in\{0,1\}^{d_{\bm{c}}}  ,
\end{equation}
where \(d_{\varphi}\) is an integer and \(I_1,\dots,I_{d_{\varphi}} \subset \llbracket 1,n\rrbracket\) are such that \(I_i  \not\subset I_{i'}\) for all \(i,i' \in \llbracket1,d_{\varphi}\rrbracket\) with \(i\neq i'\).
\end{theorem}

If a system has \(d_{\bm{c}}\) components we now consider that \textcolor{changes}{the PDMP mode $m \in \mathbb{M}$ can be converted into a multidimensional binary variable \(b_m = (c_1,\dots,c_{d_{\bm{c}}}) \in \{0,1\}^{d_{\bm{c}}}\) where for \(j \in \llbracket 1,d_{\bm{c}}\rrbracket\), \(c_j\) is 0 if the \(j\)-th component is broken and 1 otherwise}. 

\begin{definition}[Structure function \(\varphi\)]
We call the structure function of the system the Boolean function \(\varphi\) which associates 1 to the modes for which the system works and 0 to the other ones:
\textcolor{changes}{
\begin{equation}
    \varphi : b_m \longmapsto 1- \mathds{1}_{m\in\mathbb{M}_{D}}   .
\end{equation}}
\end{definition}
\begin{definition}[Coherent system]
A system is said to be coherent if:
\begin{itemize}
    \item it works when none of its components is down, i.e. \(\varphi\left((1,...,1)\right) = 1\).
    \item it does not work when all its components are broken, i.e. \(\varphi\left((0,...,0)\right) = 0\).
    \item if it does not work in a given mode then it does not work with the additional failure of a component and conversely if it works in a given mode then it still works if broken components are repaired, i.e. \(\varphi\) is non-decreasing.
\end{itemize}
\end{definition} 

\paragraph{MPS decomposition} We assume from now that the system is coherent, so we can apply \cref{th:DNF} to its structure function \(\varphi\). There are then \(d_{\text{MPS}}\) sets of components \(I_1,\dots,I_{d_{\text{MPS}}} \subset \llbracket 1,d_{\bm{c}}\rrbracket\) such that none of them is contained in another and such that the operation of all the components of a set causes the operation of the system independently of the status of the components in the other sets. These sets are the unique minimal paths sets of the system. By switching to the complementary with \(\overline{c} = 1-c\) for any \(c \in \{0,1\}\), we can notice that:
\begin{equation}
\textcolor{changes}{    \varphi(b_m) = \bigvee_{i=1}^{d_{\text{MPS}}} \bigwedge_{j\in I_i} c_j \quad \Longleftrightarrow \quad \overline{\varphi(b_m)} = \bigwedge_{i=1}^{d_{\text{MPS}}} \bigvee_{j\in I_i} \overline{c_j} \ . } 
\end{equation}
Therefore the minimal paths sets can be alternatively defined such that the failure of at least one component in each set causes the system failure. \\

\paragraph{MCS decomposition} Let us first note that since \(\varphi\) is a non-decreasing Boolean function, the function \textcolor{changes}{\(b_m\mapsto \overline{\varphi\left(\overline{b_m}\right)}\)} is also non-decreasing (it is called the dual function of \(\varphi\)). Indeed: let \textcolor{changes}{\(b_m\leq b'_m\in \{0,1\}^{d_{\bm{c}}}\)}, we have \textcolor{changes}{\(\overline{b_m}\geq\overline{b'_m}\)} so \textcolor{changes}{\(\varphi\left(\overline{b_m}\right)\geq\varphi\left(\overline{b'_m}\right)\)} and finally \textcolor{changes}{\(\overline{\varphi\left(\overline{b_m}\right)}\leq\overline{\varphi\left(\overline{b'_m}\right)}\)}. We can therefore apply \cref{th:DNF} to it and get a set of \(d_{\text{MCS}}\) lists of indices \(J_1,\dots,J_{d_{\text{MCS}}} \subset \llbracket 1,d_{\bm{c}}\rrbracket\) with \(J_i \not\subset J_{i'}\) for any \(i,i' \in \llbracket 1, d_{\text{MCS}}\rrbracket\) with \(i\neq i'\) such that:
\begin{equation}
    \textcolor{changes}{\overline{\varphi\left(\overline{b_m}\right)} = \bigvee_{i=1}^{d_{\text{MCS}}} \bigwedge_{j\in J_i} c_j 
    \quad \text{i.e} \quad  1-\varphi\left(b_m\right) = \overline{\varphi\left(b_m\right)} = \bigvee_{i=1}^{d_{\text{MCS}}} \bigwedge_{j\in J_i} \overline{c_j} \ . }
\end{equation}
These \(d_{\text{MCS}}\) sets of components \(J_1,\dots,J_{d_{\text{MCS}}}\) are the unique minimal cuts sets of the system. If all the components of a MCS are broken then this causes (from the static point of view) the system failure. And again by switching to the complementary, if at least one component per MCS works, the system failure is prevented:
\begin{equation}
   \textcolor{changes}{ \overline{\varphi(b_m)} = \bigvee_{i=1}^{d_{\text{MCS}}} \bigwedge_{j\in J_i} \overline{c_j} \quad \Longleftrightarrow \quad \varphi(b_m) = \bigwedge_{i=1}^{d_{\text{MCS}}} \bigvee_{j\in J_i} c_j \ . }
\end{equation}

\color{black}

\subsection{Parametric families of importance functions}

We below propose 3 families of IFs \(\big(\Ua{(BC)}\big)_{\tT}\), \(\big(\Ua{(MPS)}\big)_{\tT}\) and \(\big(\Ua{(MCS)}\big)_{\tT}\) to approximate the committor function \(\Uopt\), each based on a different reliability idea and parameterized by a vector \(\tT \subset \mathbb{R}^{d_{\bm{\theta}}}\) for some \(d_{\bm{\theta}}\geq 1\). To each of these families of approximations of the committor function corresponds a family \(\big(\qtheta\big)_{\tT}\) of instrumental distributions obtained by replacing \(\Uopt\) by \(\Utheta\) in \cref{eq:OptIntKer}.

One comment: since our IF only appears in the intensities and jump kernels as the ratio \(U^-_{\bm{\theta}}/\Utheta\), we only need to approximate a function proportional to \(\Uopt\) and not \(\Uopt\) itself. This means that unlike \(\Uopt\), \(\Utheta\) does not have to be interpreted as a probability and can therefore be larger than 1. On the other hand, we impose \(U_{\bm{\theta}=\bm{0}} \equiv 1\) to ensure that \(q_{\bm{0}} = \pi_{0}\) the unbiased distribution and \(\Utheta\) must be increasing in each coordinate of \(\bm{\theta}\). The intuition for the practitioner should be clear: the "larger" \(\bm{\theta}\) is, the more we bias and the more we push the trajectory to \(\mathcal{D}\) but the more unstable the likelihood ratio \(\pi_{0}/\qtheta\) could become. One just have to find a suitable trade-off. \\

\paragraph{Broken components based importance function (BC-IF)}
A natural idea is to design \(\Utheta\) as an increasing function in the number of broken components. We borrow and extend an idea from \cite{chraibiOptimalImportanceProcess2019}. Let \(d_{\bm{c}}\) be the total number of components in the system, \(\beta^{(\text{BC})}(z)\) the number of broken components when the process is in the state \(z \in E\), and \(\bm{\theta} = (\theta_1,\dots,\theta_{d_{\bm{c}}}) \in \mathbb{R}^{d_{\bm{c}}}\). We propose the following IF:
\begin{equation}
\label{eq:U_BC}
    \Ua{(BC)}(z) \, := \,\exp\left[\left(\sum_{j=1}^{d_{\bm{c}}} \theta_j \, \mathds{1}_{\beta^{(\text{BC})}(z)\geq j}\right)^2 \right] = \exp\left[\left(\sum_{j=1}^{\beta^{(\text{BC})}(z)} \theta_j \right)^2 \right]   .
\end{equation}
\noindent Theoretical arguments in favor of this particular function are given in \cite{chraibiOptimalImportanceProcess2019} for a one-dimensional form: \(\Utheta^{(\text{BC})}(z) = e^{(\theta_1 b_z)^2}\) (equivalent to \cref{eq:U_BC} by imposing \(\theta_1=\dots=\theta_{d_{\bm{c}}}\)). The intuition is that a good parametric form should favor each new component failure more and more strongly, because it is less and less likely that additional components will break without the previous ones being repaired. The form \(x\mapsto \exp(x^2)\) then guarantees that the ratios \(\sfrac{\Utheta^-}{\Utheta} \) are strictly increasing in \(\beta^{\text{(BC)}}\).

The authors of \cite{chraibiOptimalImportanceProcess2019} have successfully tested this one-dimensional approach on a parallel system with 3 components and a single failing mode (when all components are broken). It is reasonable to think that the number of components is not the core of the problem for parallel systems and that our \(d_{\bm{c}}\)-dimensional extension adds enough flexibility if needed for very large parallel systems. On the other hand, there is concern that this approach is too naive to approximate \(\Uopt\) in the case of systems with both many components and multiple failing modes. Typical failures do not involve a large number of components but a small number of key components. Using \(\Utheta^{(\text{BC})}\) produces trajectories where blindly selected components will break. With small \(\bm{\theta}\) it may not break the right components, and with large \(\bm{\theta}\) it may break too many and produce implausible trajectories (with a low likelihood ratio in the IS estimator \cref{eq:estimIS}). We must therefore take into account the role played by each component within the system. \\

\paragraph{MPS based importance function (MPS-IF)}
A MPS is said damaged when at least one of its components is broken. It is necessary that all minimal paths sets are damaged to reach system failure. For \(\bm{\theta} = \left(\theta_1,\dots,\theta_{d_{\text{MPS}}}\right) \in \mathbb{R}^{d_{\text{MPS}}}\), we recycle the idea expressed in \cref{eq:U_BC} but with this time an increasing function in \(\beta^{(\text{MPS})}(z)\) the number of damaged MPS in state \(z\):
\begin{equation}
\label{eq:U_MPS}
    \Ua{(MPS)}(z) \, := \,\exp\left[\left(\sum_{i=1}^{d_{\text{MPS}}} \theta_i \, \mathds{1}_{\beta^{(\text{MPS})}(z)\geq i}\right)^2 \right] = \exp\left[\left(\sum_{j=1}^{\beta^{(\text{MPS})}(z)} \theta_j \right)^2 \right] .
\end{equation}
The process simulated from \(\Ua{(MPS)}\) therefore prioritizes the failure of components involved in a large number of still undamaged MPS (and conversely will prioritize preventing the repair of components involved in a large number of already damaged MPS). 

The MPS-IF does not consider the number of broken components per MPS but only the presence of at least one broken component. This means that the form \(\Utheta^{\text{(MPS)}}\) does not look for breaking two components in a single MPS if they are not present in other MPS. Yet, even if this second broken component is not necessary to achieve system failure, it prevents the system from being safe in the case when the first component is repaired. Describing the functioning of the system with MCS rather than MPS leads us to a more flexible parametric family since this time we have to look at the proportion of broken components in each MCS. \\

\paragraph{MCS based importance function (MCS-IF)}
For any state \(z\in E\) and any \(i\in \llbracket 1,d_{\text{MCS}}\rrbracket\), we define \(\beta^{(\text{MCS})}_{i}(z)\) the proportion of broken components in the \(i\)-th MCS when the process is in the state \(z\). We rank these proportions in descending order and we also define  \(\beta^{(\text{MCS})}_{(i)}(z)\) the \(i\)-th largest proportion of broken components among all MCS in the state \(z\):
\begin{equation}
\label{eq:U_MCS}
    \Ua{(MCS)}(z) \, := \,\exp\left[\left(\sum_{i=1}^{d_{\text{MCS}}} \theta_i \, \beta^{(\text{MCS})}_{(i)}(z) \right)^2\right] .
\end{equation}

Here again, the process simulated from \(\Ua{(MCS)}\) prioritizes the failure of components involved in a large number of MCS or in MCS of small size (to bring the proportions of broken components closer to 1 more quickly). \\

\paragraph{Dimension reduction}
A key point for the success of the method is to get a flexible but reasonably sized family.
One way to reduce the dimension of \(\bm{\theta}\) is to impose the equality of some groups of coordinates (by ordered packets). For example, to get a vector \(\bm{\theta} \in \mathbb{R}^{d_{\bm{\theta}}}\) from \(\widetilde{\bm{\theta}} \in \mathbb{R}^{d_{\widetilde{\bm{\theta}}}}\) with \(d_{\bm{\theta}} = k\times d_{\widetilde{\bm{\theta}}}\) we impose \(\theta_i = \widetilde{\theta}_{\lfloor\frac{i-1}{k} \rfloor + 1}\) for \(i \in \llbracket 1,d_{\bm{\theta}}\rrbracket\).
This is why by choosing \(\mathds{1}_{\beta\geq i}\) in \cref{eq:U_BC} and \cref{eq:U_MPS} instead of simply  \(\mathds{1}_{\beta = i}\), and \(\beta^{(\text{MCS})}_{(i)}\) instead of \(\beta^{(\text{MCS})}_{i}\) in \cref{eq:U_MCS}, we keep a consistent (and increasing in each coordinate) \(\Utheta\) even with \(d_{\widetilde{\bm{\theta}}}=1\).  
 
\section{Algorithm and asymptotic properties}
\label{sec:Algo}

To each parametric family \(\left(\Utheta\right)_{\tT}\) of IFs designed to approximate \(\Uopt\) corresponds a family of instrumental distributions \(\left(\qtheta\right)_{\tT} \equiv \left(\pi_{\lambdatheta,\Ktheta}\right)_{\tT}\) designed to approximate \(\qopt\). Each of these distributions is defined by the jump intensity \(\lambdatheta\) and density jump kernel \(\Ktheta\) given in \cref{eq:thetaIntKer}.

The AIS method by cross-entropy that we present allows us to jointly determine a good candidate within the family \(\left(\qtheta \right)_{\tT}\) and to estimate  the probability \(P\) of the rare event that interests us. 

\subsection{Estimation procedure}
\label{subsec:Estim}

We follow the cross-entropy minimization principle \cite{deboerTutorialCrossEntropyMethod2005}. We are looking for a candidate within the family \(\left(\qtheta\right)_{\tT}\) as close as possible to the target distribution \(\qopt\) in the sense of the Kullback-Leibler divergence:
\begin{align}
\nonumber
   \underset{\bm{\theta} \in \Theta}{\arg\min} \, \mathcal{D}_{\text{KL}}\left(\qopt\|\qtheta\right) &= \underset{\bm{\theta} \in \Theta}{\arg\min} \, \, \mathbb{E}_{\qopt}\left[\log\left(\dfrac{\qopt(\mathcal{Z})}{\qtheta(\mathcal{Z})}\right)\right]  \\
   &= \underset{\bm{\theta} \in \Theta}{\arg\min} \, \int -\log\left(\qtheta(\mathcal{Z})\right) \frac{\mathds{1}_{\mathcal{Z}\in\mathcal{D}} \, \pi_{0}(\mathcal{Z})}{P} d\zeta(\mathcal{Z}) \nonumber \\ 
   &=  \underset{\bm{\theta} \in \Theta}{\arg\min} \, \, \mathbb{E}_{\pi_{0}}\left[-\mathds{1}_{\mathcal{Z}\in\mathcal{D}}\,\log\left(\qtheta\left(\mathcal{Z}\right)\right)\right]  .
\end{align}

\paragraph{Sequential algorithm} The function \(\bm{\theta} \mapsto \mathbb{E}_{\pi_{0}}\left[-\mathds{1}_{\mathcal{Z}\in\mathcal{D}}\,\log\left(\qtheta\left(\mathcal{Z}\right)\right)\right]\) is estimated by importance sampling under an initial instrumental distribution \(q_{\bm{\theta}^{(\bm{1})}}\), we determine \(\bm{\theta}^{(2)}\) which minimizes this estimate and we repeat the scheme. To save the simulation budget, we reuse at each iteration all the trajectories already drawn to perform the minimization step. Similarly, all the trajectories generated during the algorithm are recycled to produce the final estimator of \(P\). To summarize, at iteration \(\ell\in\mathbb{N}^*\):
\begin{enumerate}
    \item \textbf{Simulation phase.} Generate \(n_{\ell}\) trajectories \(\mathcal{Z}_1^{(\ell)},\dots,\mathcal{Z}_{n_{\ell}}^{(\ell)}  \, \overset{\text{i.i.d.}}{\sim} \,q_{\bm{\theta}^{(\ell)}}\).

    \item \textbf{Optimization phase.} Update the parameter of the instrumental distribution with the \(\ell\) last samples drawn  \(\left(\mathcal{Z}_k^{(1)}\right)_{k=1}^{n_1},\dots,\left(\mathcal{Z}_k^{(\ell)}\right)_{k=1}^{n_{\ell}}\):
    \begin{equation}
    \label{eq:MinCE}
         \bm{\theta}^{(\ell+1)} =  \underset{\bm{\theta}\in\Theta}{\arg\min } \left\{ - \sum_{r=1}^{\ell} \sum_{k=1}^{n_r} \mathds{1}_{\mathcal{Z}_k^{(r)}\in\mathcal{D}} \frac{\pi_{0}\big(\mathcal{Z}_k^{(r)}\big)}{q_{\bm{\theta}^{(r)}}\big(\mathcal{Z}_k^{(r)}\big)} \log \left[\qtheta \big(\mathcal{Z}_k^{(r)}\big)\right]\right\} .
    \end{equation}  
\end{enumerate}
\textbf{Estimation phase} at the final iteration \(L\) (with \(N_L = \sum_{\ell=1}^L n_{\ell}\) the total budget), we reuse all past samples to get the final estimator of \(P\)~:
\begin{equation}
    \label{eq:EstimateP}
    \widehat{P}_{N_L} = \dfrac{1}{N_{L}} \sum_{\ell=1}^L \sum_{k=1}^{n_{\ell}} \mathds{1}_{\mathcal{Z}_k^{(\ell)}\in\mathcal{D}} \frac{\pi_{0}\big(\mathcal{Z}_k^{(\ell)}\big)}{q_{\bm{\theta}^{(\ell)}}\big(\mathcal{Z}_k^{(\ell)}\big)} .
\end{equation}

\subsection{Asymptotic optimality and confidence interval}
\label{subsec:Asymp}
Using theorems 2 and 3 from \cite{delyonAsymptoticOptimalityAdaptive2018}, we can determine sufficient criteria to ensure the consistency and asymptotic normality of the estimator~\cref{eq:EstimateP}.

\begin{hypothesis}[Assumptions on the PDMP]
\label{hyp:PDMP}
The PDMP of distribution \(\pi_{0}\) with states in \(E\), jump intensity \(\lambdazero\) and jump kernel \(\Kzero\) verifies the following conditions:
\begin{enumerate}
    \item There exist \(\lambda_{\min},\lambda_{\max}>0\) such that for any \(z \in \text{supp}(\lambda)\), \(\lambda_{\min}\leq \lambdazero(z) \leq \lambda_{\max}\).

    \item There exist \(K_{\min},K_{\max}>0\) such that for any \(z^-\in E\) and \(z \in \text{supp} \left(\Kzero\left(z^-,\cdot \right)\right)\), \( K_{\min}\leq \Kzero(z^-,z) \leq K_{\max}\).
  
    \item There exists \(t_{\varepsilon}>0\) such that for any \(z^- \in \partial E\) and \(z \in \text{supp} \left(\Kzero\left(z^-,\cdot \right)\right)\), \(t_z^{\partial} \geq t_{\varepsilon}\).

\end{enumerate}
\end{hypothesis}

\begin{hypothesis}[Assumptions on the parametric family \(\left(\Utheta\right)_{\tT}\) and on the set \(\Theta\)]
\label{hyp:U_theta}
The set of parameters \(\Theta\) and the parametric family \(\left(\Utheta\right)_{\bm{\theta}\in\Theta}\) of IFs verify the following conditions:

\begin{enumerate}
    \item \(\Theta\) is a compact subset of  \(\mathbb{R}^{d_{\bm{\theta}}}\) for some \(d_{\bm{\theta}}>0\).
    
    \item \(\thetaopt \in \Theta\) is the unique minimizer of \(\bm{\theta} \mapsto  \mathbb{E}_{\pi_{0}}\left[-\mathds{1}_{\mathcal{Z}\in\mathcal{D}}\,\log\left(\qtheta\left(\mathcal{Z}\right)\right)\right]\).
    \item There exist \(U_{\min},U_{\max} > 0\) such that for any \(\bm{\theta}\in \Theta\) and \(z\in E\), \(U_{\min}\leq \Utheta(z) \leq U_{\max}\).

\end{enumerate}
\end{hypothesis}

\begin{theorem}
\label{th:Asymp}
Under \cref{hyp:PDMP} and \cref{hyp:U_theta}, with \(V : \bm{\theta} \mapsto \mathbb{E}_{\pi_{0}}\left[\mathds{1}_{\mathcal{Z}\in\mathcal{D}} \frac{\pi_{0}(\mathcal{Z})}{\qtheta(\mathcal{Z})} \right] - P^2\), we have
\begin{equation}
\label{eq:Asympt}
    \bm{\theta}^{(L)}\, \overset{\text{a.s.}}{\longrightarrow}\, \thetaopt \quad \text{and} \quad \sqrt{N_L}\left(\widehat{P}_{N_L} - P\right) \, \overset{\mathcal{L}}{\longrightarrow} \, \mathcal{N}\left(0,V\left( \thetaopt\right)\right)
\end{equation}
if one of the two following conditions are satisfied:
\begin{enumerate}
    \item \(n_{\ell}>0\) for any \(\ell>0\) and \(L\rightarrow + \infty\),
    \item \(L<\infty\), \(n_{L-1} \rightarrow + \infty\) and \(n_L / n_{L-1} \rightarrow +\infty\). 
\end{enumerate}

\end{theorem}

The asymptotics can therefore be taken either in the number of iterations \(L\) or in the size of the last two samples \(n_{L-1}\) and \(n_{L}\). These are two different yet specific ways to make the total number of simulated trajectories tend towards infinity. The first case, more standard, is proven in \cref{subsec:Proof}. In the second case, with a fixed number of iterations \(L\), if the number of simulated trajectories at the second to last iteration \(n_{L-1}\) tends to infinity, then we minimize \(\bm{\theta} \mapsto \mathbb{E}_{\pi_{0}}\left[-\mathds{1}_{\mathcal{Z}\in\mathcal{D}}\,\log\left(\qtheta\left(\mathcal{Z}\right)\right)\right]\) which gives $\thetaopt$. Thus at the last iteration \(L\), the trajectories are generated according to \(q_{\bm{\theta}_{\text{opt}}}\). It is then sufficient that \(n_L/n_{L-1}\) tends to infinity for the proportion of trajectories drawn according to \(q_{\bm{\theta}_{\text{opt}}}\) to converge to one.

We can also propose a consistent estimator of the asymptotic variance \(V(\bm{\theta}_{\text{opt}})\):
\begin{equation}
\label{eq:Var}
   \widehat{\sigma}^2_{N_{L}} = \dfrac{1}{N_{L}} \sum_{\ell=1}^L \sum_{k=1}^{n_{\ell}} \mathds{1}_{\mathcal{Z}_k^{(\ell)}\in\mathcal{D}} \frac{\pi_{0}\big(\mathcal{Z}_k^{(\ell)}\big)^2}{q_{\bm{\theta}^{(\ell)}}\big(\mathcal{Z}_k^{(\ell)}\big)^2} - \widehat{P}_{N_{L}}^2 .
\end{equation}
It follows that an asymptotic confidence interval for \(P\) with the conditions of \cref{th:Asymp} is given by:
\begin{equation}
\label{eq:IC}
   \mathbb{P} \left(P \in \left[\widehat{P}_{N_L} - v_{1-\alpha/2}\, \widehat{\sigma}_{N_L} \,
   N_L^{-1/2} ~ ; ~ \widehat{P}_{N_L} + v_{1-\alpha/2}\, \widehat{\sigma}_{N_L} \, N_L^{-1/2} \right]\right) \longrightarrow  1-a ,
\end{equation}
where \(v_{1-\alpha/2}\) is the (\(1-\alpha/2\))-quantile of the \({\cal N}(0,1)\) distribution. \\

\section{Implementation guidelines}
\label{sec:Implement}
We discuss here the implementation of our method. 

\paragraph{Initialization of the algorithm} It is known that the choice of the initial distribution of an AIS method is crucial. A common option that has the advantage of being without a priori is to take \(q_{\bm{\theta}^{(0)}} = \pi_{0}\), but it is not suitable to deal with rare events. Let us recall that we would like to minimize \(\mathbb{E}_{\pi_{0}}\left[\mathds{1}_{Z\in\mathcal{D}} \log \qtheta(Z) \right]\) but we minimize in practice an empirical approximation \cref{eq:MinCE}. With a poorly chosen initial auxiliary distribution, the minimizer of the approximation could be too far from the true minimizer. It is then difficult to find the right track over the iterations and the final result of the procedure depends therefore a lot on this initial choice. 

In the starting configuration of our system, the first spontaneous jump can only be the failure of a component since none are broken at the initial time. We are therefore able to set a time limit \(\widetilde{t}\) and a threshold probability \(\widetilde{p}\) and to determine the smallest \(\widetilde{\theta}\) (of dimension 1) such that the probability under \(\pi_{\widetilde{\theta}}\) that the first spontaneous jump takes place before the time \(\widetilde{t}\) is larger than \(\widetilde{p}\). We can then start the Cross-Entropy with \(\bm{\theta}^{(0)} = \left(\widetilde{\theta},\dots,\widetilde{\theta} \right)\) and
\begin{equation}
\label{eq:thetaInitCE}
    \widetilde{\theta} :=  \inf \left\{\theta \in \mathbb{R}_+ \, : \,\mathbb{P}_{q_{\theta}}\left(T_{z_0} \leq \widetilde{t} \mid Z_0 = z_0\right) \geq \widetilde{p}\right\}  ,
\end{equation}
with \(T_{z_0}\) the time of the first random jump and hence the time of the first component failure. \\

\paragraph{Sampling size policy and stopping criterion}
Given a fixed simulation budget, we would like to choose \(n_{\ell}\) the sample size at iteration \(q\), and \(L\) the total number of iterations. Let us recall that \(N_L = \sum_{\ell=1}^L n_{\ell}\) is the total number of trajectories to be drawn. 

If we do not opt for a recycling scheme, it is imperative to set \(n_{\ell}\) large enough (at the very least 100) even if it means doing few iterations, because we cannot rely on past samples to approximate the true objective function in \cref{eq:MinCE}. On the other hand with a recycling scheme, we would ideally like to choose \(n_{\ell}\) as small as possible in order to perform as many iterations as possible, thus as many minimizations as possible, and to give ourselves the best chances to get close to \(\qopt\). 

In practice, it all depends on the optimization method used to solve \cref{eq:MinCE}. If it is expensive, we cannot afford too many iterations. Moreover, the cost of calling the function to be minimized depends linearly on the number of terms in the sum, so the minimization will be more and more expensive with each iteration (using a recycling scheme). To get an idea, if we do not want the cost dedicated to the optimization to exceed the cost dedicated to the simulation, the number of iterations \(L\) should be smaller than \(2 c_S/c_O - 1\) with \(c_S\) the computational cost of simulating one trajectory and \(c_O\) the computational cost of minimizing \cref{eq:MinCE} for a single trajectory. 

For a fixed simulation budget \(N_L\), we propose to determine \(n_{\ell}\) at iteration \(q\) as the minimum between the remaining budget \(N_L - N_{q-1}\) and the (random) smallest number of trajectories to be drawn such that \(n_{\text{CE}}\) of them belong to \(\mathcal{D}\) (with for example \(n_{\text{CE}} = 10\) in the case of a total budget \(N_L = 10^4\)). \\

\paragraph{Numerical optimization} It is necessary at each iteration to call an optimization program to solve \cref{eq:MinCE}. Let us point out that with a large number of iterations and few new trajectories at each sample, it is not useful to determine the true minimizer (which would imply using sophisticated methods adapted to non-convex problems). We only need to improve our instrumental distribution a little bit at each iteration. 

We employed the BFGS method \cite{daiConvergencePropertiesBFGS2002} that can be found in many  toolboxes. We used the function \texttt{minimize(\(\cdot\),method=BFGS)} from the \texttt{scipy.optimize} toolbox in Python \cite{virtanen2020scipy}. This function performs better when given the explicit gradient of the function to be minimized rather than letting it approximate it by finite differences. We give in \cref{subsec:Grad} the explicit gradient of the function \(\bm{\theta}\mapsto \log \qtheta(\cdot)\) involved in \cref{eq:MinCE}. 

One last advice: a classical stopping criterion of a BFGS method is to obtain a sufficiently small gradient (i.e. close to zero). The default threshold in \texttt{scipy.optimize} is \(10^{-5}\). It is better to drastically lower this threshold to \(10^{-20}\) for example because since the set of possible trajectories is a very high dimensional space, the densities of the trajectories are very small and the gradient of the objective function is small. \\

\paragraph{PDMP simulation}
\label{par:Simu}
 It is assumed that a suitable numerical code can be called up to compute the flow. It represents the main computational cost of the simulation.
 There are several methods, exact or approximate, to simulate the jump times of a PDMP \cite{lemaireExactSimulationJump2018,bouissouEfficientMonteCarlo2014,veltzNewTwistSimulation2015,lemaireThinningMultilevelMonte2018,riedlerAlmostSureConvergence2013}. Since our main constraint comes from the computation of the flow, we will opt for a method that is sparing in the number of calls to the flow. We adapt for this purpose the algorithm 3.4 of \cite{desaportaNumericalMethodsSimulation2015} which is intended for the case where the flow is explicitly known. This algorithm is based on a thinning principle which is usual for simulating time inhomogeneous Poisson processes \cite{lewisSimulationNonhomogeneousPoisson1979}. The \texttt{PyCATSHOO} toolbox \cite{chraibiPyCATSHOONewPlatform2016} is an EDF-developed computer code that enables such simulations.\\

\paragraph{MPS/MCS decomposition} Listing all the MPS or MCS of a system is an NP-hard problem. This task can be done by hand on the SFP system that we present in \cref{subsec:PrezPDMP} but it becomes impractical in the case of very large, highly redundant systems. The literature presents more methods to determine the MCS than the MPS of a system but the two problems are strictly equivalent. Fault trees are the most common representation of systems in the static Boolean approach and the search for the MCS of the system belongs to the field called fault tree analysis (FTA) (see \cite{ruijtersFaultTreeAnalysis2015} for a recent survey). This is an old  but still active field in the industrial and academic communities. New approaches based on the differential logic calculus offer other perspectives on the decomposition of the structure function \cite{rusnakLogicDifferentialCalculus2022}.\\

\section{Numerical experiments}
\label{sec:Exp}
In this section we present the results obtained with our method, first on series and parallel systems, and then on the spent fuel pool system represented in \cref{fig:SFP_Representation}. We compare the performances of the AIS method with each of the three families of IFs (BC-IF, MPS-IF and MCS-IF) to a CMC method. 

\subsection{Series and parallel systems}

We study series and parallel systems with \(d_{\bm{c}}\) components. \textcolor{changes}{We set $\mathbb{M} = \{0,1\}^{d_{\bm{c}}}$. The mode of the system is $m = \left(m^{(1)},\dots,m^{(d_{\bm{c}})}\right)$ where for \(j \in \{1,\dots,d_{\bm{c}}\}\), the status of the $j$-th component $m^{(j)} = 1$ if the component is active and 0 if it is broken.} 
The mode at time \(t \geq 0\) thus corresponds to the current status of each component: \(M_t = (M^{(1)}_t,\dots,M^{(d_{\bm{c}})}_t)\). Therefore in the case of series systems \(M_t \in \mathbb{M}_{D}\) if there is  \(j \in \{1,\dots,d_{\bm{c}}\}\) such that \(M_t^{(j)} = 0\), and in the case of parallel systems \(M_t \in \mathbb{M}_{D}\) if \(M_t^{(j)} = 0\) for any \(j \in \{1,\dots,d_{\bm{c}}\}\).

Under distribution \(\pi_{0}\), for \(j \in \{1,\dots,d_{\bm{c}}\}\), the $j$-th component has a jump rate \(\lambda^{(j)}_{0}\) that depends on its status (in other words it has a constant failure rate and a constant repair rate). From the state \(z^-\), the next jump occurs at a random time of jump intensity \(\lambdazero(z^-) = \sum_{j=1}^{d_{\bm{c}}} \lambda^{(j)}_{0}(z^-)\). At each jump from state \(z^-\), only one component is randomly selected with probability \(\lambda^{(j)}_{0}(z^-)/\lambdazero(z^-)\) for \(j \in \{1,\dots,d_{\bm{c}}\}\) and it then changes status.  

\textcolor{changes}{The system failure is reached either as soon as the process has spent a total time larger than \(x^{(1)}_{\max}\) in \(\mathbb{M}_D\) (global grace period), or when it remains in \(\mathbb{M}_D\) a time larger than \(x^{(2)}_{\max}\) without leaving it (local grace period). We note \(X_t = (X^{(1)}_{t},X^{(2)}_{t},X^{(3)}_{t})\) the position of the process at time \(t\geq 0\) with \(X^{(1)}_{t}\) the total time spent in \(\mathbb{M}_D\) during the entire trajectory, \(X^{(2)}_{t}\) the elapsed time since the entry in \(\mathbb{M}_D\) if the process is there and 0 otherwise, and finally $X^{(3)}_{t} = t$ the total elapsed time. For an initial time \(t_0>0\) and a departure state \(Z_{t_0} = \left(X_{t_0},M_{t_0}\right) \), the flow of the PDMP is given by $\Phi_{Z_{t_0}} : h \mapsto \left(X_{t_0+h},M_{t_0+h}\right)$ with~:
\begin{align}
X^{(1)}_{t_0+h} &= X^{(1)}_{t_0} \mathds{1}_{M_{t_0} \notin \mathbb{M}_D}  + \left(X^{(1)}_{t_0}+h\right) \mathds{1}_{M_{t_0} \in \mathbb{M}_D}  , \\
X^{(2)}_{t_0+h} &= \left(X^{(2)}_{t_0}+h\right) \, \mathds{1}_{M_{t_0} \in \mathbb{M}_D} , \\
X^{(3)}_{t_0+h} &= t_0 + h .
\end{align}
}

\paragraph{Importance distribution for series and parallel systems} In this subsection we will note \(\beta_z \equiv \beta^{\text{(BC)}}(z)\) the number of broken components in state \(z \in E\). As seen in \cref{subsec:MPSMCS}:
\begin{enumerate}
    \item A series system with \(d_{\bm{c}}\) components has 1 MPS containing all the components and \(d_{\bm{c}}\) MCS containing each 1 component. Thus in series systems \(\Ua{(MCS)} = \Ua{(BC)}\) and \(\Ua{(MPS)} = \exp\left[\theta_1^2\, \mathds{1}_{\beta_z\geq 1} \right]\). 
 \item A parallel system with \(d_{\bm{c}}\) components has \(d_{\bm{c}}\) MPS containing each 1 component and 1 MCS containing all the components. Thus in parallel systems \(\Ua{(MPS)} = \Ua{(BC)}\) and \(\Ua{(MCS)} = \exp [ (\theta_1 \beta_z/ d_{\bm{c}} )^2 ]\). 
 \end{enumerate} 

For these cases, we give in \cref{tab:jumpIntensity_SeriesParallel} explicit  expressions of the jump intensity \(\lambdatheta\) and jump kernel \(\Ktheta\) of the importance density \(\qtheta\) from the marginal jump intensities \(\left(\lambdatheta^{(j)}\right)_{j=1}^{d_{\bm{c}}}\).

\begin{table}

\begin{center}
 
\renewcommand{\arraystretch}{1.8}
{\footnotesize
\begin{tabular}{|c|c|c|} 
\cline{2-3}  
\multicolumn{1}{c|}{} & \(\lambdatheta^{(j)}\) in a \textbf{series system} & \(\lambdatheta^{(j)}\) in a \textbf{parallel system} \\  
\hline 
With \(\Ua{(BC)}\) & \multicolumn{2}{c|}{\(\lambda^{(j)}_{0}(z) \exp\left[\theta_{\beta_z+m^{(j)}}^2 + 2(2 m^{(j)} - 1) \theta_{\beta_z + m^{(j)}} \sum_{i=1}^{\beta_z} \theta_i  \right]\)} \\  
\hline 
With \(\Ua{(MPS)}\) & \(\lambda^{(j)}_{0}(z) \exp\left[\theta_1^2\left(\mathds{1}_{\beta_z\geq 2(1-m^{(j)})} - \mathds{1}_{\beta_z\geq 1}\right) \right]\) & Same as BC case \\  
\hline 
With \(\Ua{(MCS)}\) & Same as BC case & \( \lambda^{(j)}_{0}(z) \exp\left[\left(\sfrac{\theta_1}{d_{\bm{c}}}\right)^2 \left(1 + 2(2 m^{(j)} - 1) \beta_z \right) \right]\) \\
\hline
\end{tabular}
}
\caption{\footnotesize\textbf{Marginal importance jump intensity of the $j$-th component for the importance process on series and parallel systems.} The jump intensity \(\lambdatheta\) and the jump kernel density \(\Ktheta\) of the importance distribution \(\qtheta\) are simply written: \(\lambdatheta(z) = \sum_{j=1}^{d_{\bm{c}}} \lambdatheta^{(j)}(z)\) and \(\Ktheta\left(z,z^{(j)}\right) = \lambdatheta^{(j)}/\lambdatheta(z)\) with \(z^{(j)}\) the same state as \(z\) except for \(m^{(j)}\) the status of the $j$-th component.}\label{tab:jumpIntensity_SeriesParallel}
\end{center}
\end{table}

Each of the two systems presents a different challenge for importance sampling. The series system requires multi-modal importance distribution since the failure can come from any component, and the importance distribution for a parallel system must produce sequences where all components fail that are plausible from the perspective of jump times. \\

\paragraph{Results}

We compare on a series and on a parallel system the performance of a CMC method with a sample size ranging from \(10^5\) to \(10^7\) to our three versions of the AIS method corresponding to the three families of approximations of the committor function with a sample size ranging from \(10^3\) to \(10^4\). For both the series and parallel systems, we generated trajectories of duration \(t_{\max} = 1500\) with global grace period \(x^{(1)}_{\max} = 75\) and local grace period \(x^{(2)}_{\min} = 50\). Each system has five components. The jump parameters of the two systems are described in \cref{tab:seriesParallelParameters}. The results obtained on the series system, resp. parallel system, are described in \cref{tab:results_SeriesSystem}, resp. in \cref{tab:results_ParallelSystem}.

\begin{table}

\begin{center}
\renewcommand{\arraystretch}{1.3}
{\footnotesize
\begin{tabular}{|c | c | c | c | c | }
\hline
\textbf{Method} & \(N_L\) &  \textcolor{changes}{\(\widehat{P}_{N_L}\)} & \textcolor{changes}{\(\widehat{\sigma}_{N_L}\)} & 95\% confidence interval  \\
\hline \hline 
 & \(10^5\) & \(4 \times 10^{-5}\) &  $6.32 \times 10^{-3}$ & \(\left[ 8.02\times 10^{-7} \, ; \, 7.92\times 10^{-5} \right]\) \\
CMC & \(10^6\) & \(2.9 \times 10^{-5}\) &  $5.38 \times 10^{-3}$ & \(\left[ 1.84 \times 10^{-5} \, ; \, 3.96\times 10^{-5} \right]\)\\
 & \(10^7\) & \(2.7 \times 10^{-5}\) &  $5.19 \times 10^{-3}$ & \(\left[ 2.38 \times 10^{-5} \, ; \, 3.02\times 10^{-5} \right]\) \\
\hline 

\multirow{2}{*}{IS with \(\Ua{(MPS)}\)} & \(10^3\) & \(2.82\times 10^{-5}\) & $2.26 \times 10^{-5}$ & \(\left[2.68 \times 10^{-5} \, ; \, 2.96 \times 10^{-5}\right]\) \\
 & \(10^4\) & \(2.91\times 10^{-5}\) & $2.24 \times 10^{-5}$ & \(\left[2.87 \times 10^{-5} \, ; \, 2.95 \times 10^{-5}\right]\)\\
 \hline
\multirow{2}{*}{IS with \(\Ua{(MCS)} = \Ua{(BC)}\)} & \(10^3\) &  \(2.96 \times 10^{-5}\) & $2.25 \times 10^{-5}$ & \(\left[ 2.82\times 10^{-5} \, ; \, 3.09\times 10^{-5} \right]\) \\
 & \(10^4\) & \(2.89 \times 10^{-5}\)
 & $2.25 \times 10^{-5}$ & \(\left[ 2.84\times 10^{-5} \, ; \, 2.93\times 10^{-5} \right]\) \\

 \hline
\end{tabular}
}
\caption{\footnotesize\label{tab:results_SeriesSystem}\textbf{Results on the series system case} (with jump rates of the five components from \cref{tab:seriesParallelParameters} in appendix). The 3 AIS versions were initialized according to the method described in \cref{sec:Implement}: with the smallest one-dimensional \(\bm{\theta}\) such that the probability that at least one component failure occurs before the end of the simulation is larger than \(1/3\). At each iteration, we generate trajectories until we have \(n_{\text{CE}} = 10\) failures before updating \(\bm{\theta}\) for \(N_L = 10^3\) and \(n_{\text{CE}} = 50\) for \(N_L = 10^4\). We stop when the total budget \(N_L\) is reached. The effective dimension of the vector \(\bm{\theta}\) is not modified. For the CMC method, we simply generate $N_L$ trajectories and we count the proportion of faulty trajectories. \textcolor{changes}{The estimated probability \({\widehat{P}_{N_L}}\) is given by \cref{eq:EstimateP} and the estimated asymptotic standard deviation \({\widehat{\sigma}_{N_L}}\) is  given by \cref{eq:Var}.}}
\end{center}
\end{table}

\begin{table}

\begin{center}
\renewcommand{\arraystretch}{1.3}
{\footnotesize
\begin{tabular}{|c | c | c | c | c | }
\hline
\textbf{Method} & \(N_L\) &  \textcolor{changes}{\(\widehat{P}_{N_L}\)} & \textcolor{changes}{\(\widehat{\sigma}_{N_L}\)} & 95\% confidence interval  \\
\hline \hline 
 & \(10^5\) & \(8 \times 10^{-5}\) &  $8.94 \times 10^{-3}$ & \(\left[ 2.46 \times 10^{-5} \, ; \, 1.35\times 10^{-4} \right]\) \\
CMC & \(10^6\) & \(6.7 \times 10^{-5}\) &  $7.54 \times 10^{-3}$ & \(\left[5.10 \times 10^{-5} \, ; \, 8.30\times 10^{-5} \right]\)\\
 & \(10^7\) & \(6.73 \times 10^{-5}\) &  $8.2 \times 10^{-3}$ & \(\left[ 6.22 \times 10^{-5} \, ; \, 7.24\times 10^{-5} \right]\) \\
\hline 
\multirow{2}{*}{IS with \(\Ua{(MPS)} = \Ua{(BC)}\)} & \(10^3\) &  \(4.85 \times 10^{-5}\) & $1.77 \times 10^{-4}$ & \(\left[ 3.76\times 10^{-5} \, ; \, 5.94\times 10^{-5} \right]\) \\
 & \(10^4\) & \(5.80 \times 10^{-5}\)
 & $2.53 \times 10^{-4}$ & \(\left[ 5.29\times 10^{-5} \, ; \, 6.31\times 10^{-5} \right]\) \\
 \hline 
\multirow{2}{*}{IS with \(\Ua{(MCS)}\)} & \(10^3\) & \(5.94\times 10^{-5}\) & $2.95 \times 10^{-4}$ & \(\left[4.12 \times 10^{-5} \, ; \, 7.77 \times 10^{-5}\right]\) \\
 & \(10^4\) & \(7.01\times 10^{-5}\) & $7.14 \times 10^{-4}$ & \(\left[5.61 \times 10^{-5} \, ; \, 8.41 \times 10^{-5}\right]\)\\

 \hline
\end{tabular}
}
\caption{\footnotesize\label{tab:results_ParallelSystem}\textbf{Results on the parallel system case} (with jump rates of the five components from \cref{tab:seriesParallelParameters} in appendix). Same notation as in Table \ref{tab:results_SeriesSystem},
except that the 3 IS forms were initialized with the smallest one-dimensional \(\bm{\theta}\) such that the probability that at least one component failure occurs before \(t_{\max}/d_{\bm{c}}\) is larger than \(1/3\). 
}
\end{center}
\end{table} 

The AIS method performs better than the CMC method  in all configurations. The estimated probabilities are of the same order and the confidence intervals produced by the AIS method for a given sample size are of comparable length to the confidence intervals produced by the CMC method for a sample size \(10^4\) larger. It can be seen that the best performance is obtained on the series system despite a slightly lower failure probability. This result is not surprising since the failed trajectories of a series system generally contain few jumps and thus produce likelihood ratios that are easier to stabilize. Since only the failure (and non-repair) of a single component is necessary for the system to fail, the MPS and BC/MCS methods have the same effectiveness here. For the parallel system on the other hand, the BC/MPS method benefits from additional degrees of freedom compared to the MCS method which seems to make a small difference at the end. In particular, the BC/MPS form allows the speed at which component failures must follow each other until the failure mode is reached to be dosed precisely.

\subsection{The spent fuel pool}
The roles of the components \(\left(c_j\right)_{j=1}^{d_{\bm{c}} = 15}\) are described in \cref{fig:SFP_Representation}. \textcolor{changes}{We set $\mathbb{M} = \{-1,0,1\}^{d_{\bm{c}}}$. The mode of the system is $m = \left(m^{(1)},\dots,m^{(d_{\bm{c}})}\right)$ where for \(j \in \{1,\dots,d_{\bm{c}}\}\), the status of the $j$-th component $m^{(j)} = 1$ if the component is active, 0 if it is inactive and -1 if it is broken.}.
 The mode \(M_t = (M^{(1)}_t,\dots,M^{(d_{\bm{c}})}_t)\) of the process at time $t\geq 0$ corresponds to the current status of each component. Recall that we have \(M_t \in \mathbb{M}_{D}\) if at time \(t\geq 0\) all MPS are damaged or equivalently if at least one MCS has all its components broken.

\textcolor{changes}{
We note \(X_t = (X^{(1)}_t,X^{(2)}_t,X^{(3)}_t )\) the position of the process  at time \(t\geq 0\) with \(X^{(1)}_t\) the temperature of the water in the pool in \(\degree \text{C}\), \(X^{(2)}_t\) the water level in the pool in meters (\(\text{m}\)) and $X^{(3)}_{t}=t$ the total elapsed time. The evolution of these variables is described by the system of ordinary differential equations:
%
%
 \begin{align}
\label{eq:Temperature}
    \dfrac{d X^{(1)}_t}{dt} &= \mathds{1}_{X^{(1)}_t<100}\times \dfrac{r + \rho C Q (x^{(1)}_{S} - X^{(1)}_t)\mathds{1}_{M_t \notin \mathbb{M}_{D}}}{\rho C A X^{(2)}_t}  , \\
\label{eq:WaterLevel}
    \dfrac{d X^{(2)}_t}{dt} &= - \mathds{1}_{X^{(1)}_t=100}\times \dfrac{r}{\rho C A \ell}  ,\\
    \dfrac{d X^{(3)}_t}{dt} &= 1 ,
\end{align}   
}%
where the physical parameters are given in the \cref{tab:physicalVar} (values taken from \cite{chraibiPyCATSHOONewPlatform2016}).  


\begin{table}

\begin{center}
\renewcommand{\arraystretch}{1.0}
{\footnotesize
\begin{tabular}{|c | c | c |  }
\hline
\textbf{Physical parameters} & Value & Description  \\
\hline \hline
\(r\) & \(2.106\times 10^{10} \, \text{J} \cdot \text{h}^{-1}\) & Residual power of the fuel \\ 
\(C\) & \(4180 \, \text{J}\cdot \text{Kg}^{-1}\cdot \degree\text{K}^{-1}\) & Mass heat capacity \\ 
\(\rho\) & \(990 \, \text{Kg}\cdot\text{m}^{-3} \) & Density of the water \\  
\(A \) & \(77 \text{m}^2\) & Area of the pool.  \\ 
\(x^{(1)}_{S}\) & \(15^ \degree \text{C}\) & Temperature of the water sources \\
\(Q\) & \(550 \, \text{m}^3 \,\text{h}^{-1} \) & The debit water \\
\(\ell\) & \(2.257 \times 10^6 \, \text{J} \cdot \text{Kg}^{-1}\) & Latent heat of vaporization \\
\(t_{\max}\) & \(3600 \, \text{h}\) & Duration of the mission \\
\(x^{(2)}_{0}\) & \(19\, \text{m}\) & Initial level of water in the pool \\
\(x^{(2)}_{\min}\) &  \(16\, \text{m}\) & Critical threshold of the level of water in the pool\\
\hline
\end{tabular}
}
\caption{\footnotesize\label{tab:physicalVar}\textbf{Physical parameters} of the SFP. Values taken from \cite{chraibiPyCATSHOONewPlatform2016}.}
\end{center}
\end{table} 

Under distribution \(\pi_{0}\), each component \(m^{(j)}\) has a jump rate \(\lambda^{(j)}_{0}\) which depends on its status and on the values of the physical variables of the system. The jump intensity of the PDMP in a state \(z\in E\) is the sum of the jump rates of the components in state \(z\): \(\lambdazero(z) = \sum_{j=1}^{d_{\bm{c}}} \lambda^{(j)}_{0}(z)\). At each jump from state \(z^-\), a component \(c_j\) is randomly selected with probability \(\lambda^{(j)}_{0}(z^-)/\lambdazero(z^-)\) and changes status (it is repaired if it was down, and fails otherwise). The system automatically reconfigures itself by enabling or disabling components so that exactly 1 MPS has all its components active if possible (be careful not to confuse inactive component \(m^{(j)} = 0\) and broken component \(m^{(j)} = - 1\)).  

It is assumed that no water can be re-injected into the pool in case of evaporation for the duration of the mission \(t_{\max}\). Once in \(\mathbb{M}_D\) there is a first grace period before the temperature of the water reaches 100\(\degree \text{C}\), but this temperature can go back down once the system is repaired. Then we have a second grace period before the water level in the pool reaches a critical threshold \(x^{(2)}_{\min}\). In our model, the evaporated water is lost and the water level cannot rise again if the system is repaired.\\

\paragraph{Results} 

\begin{figure}
\begin{center}
    \includegraphics[scale=0.23]{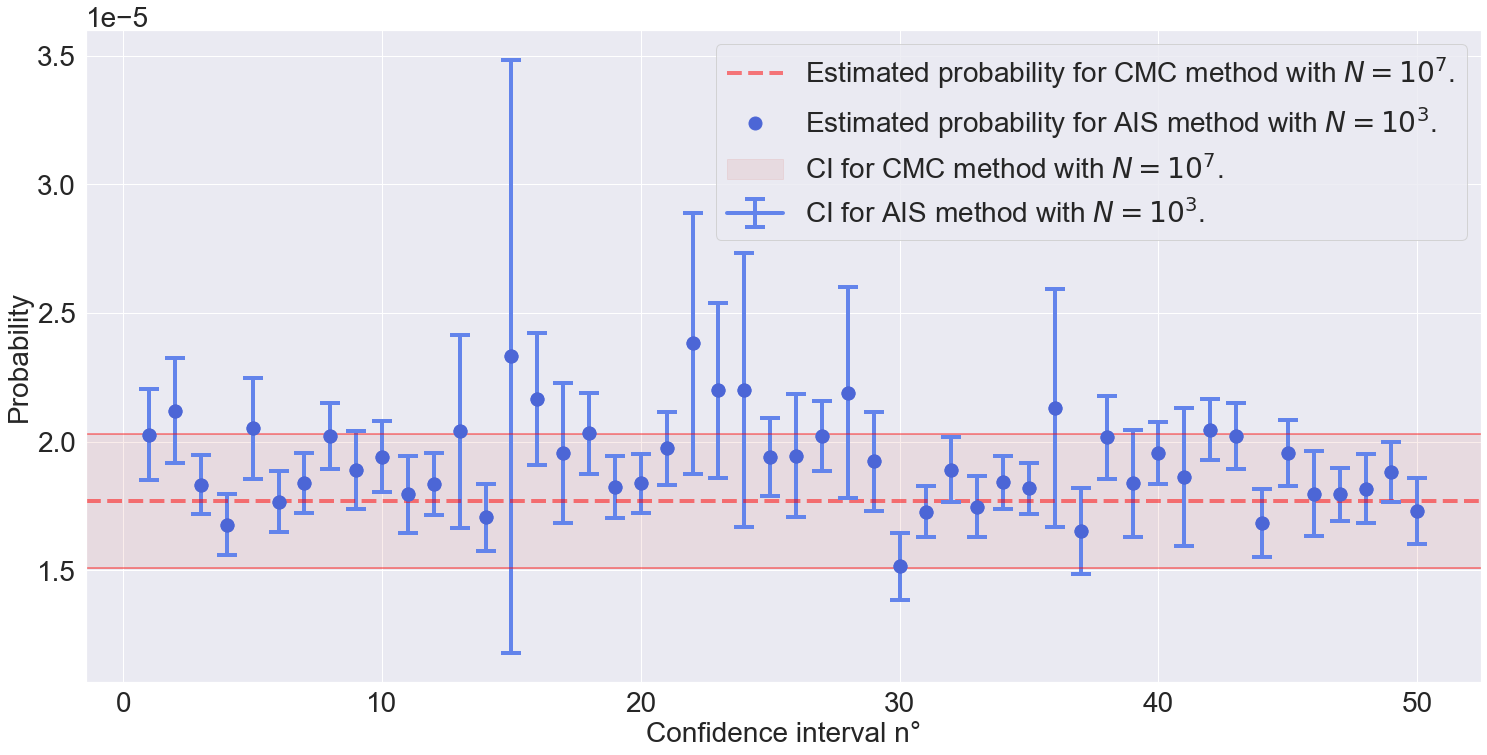}
    \caption{\small\textbf{Comparison of 50 confidence intervals at 95\% level obtained with MPS-IF approximation.} Each confidence interval corresponds to a run of the AIS MPS-IF method on the standard case of the SFP (\cref{tab:standardCase}) with \(10^3\) trajectories (same conditions as for \cref{tab:resultsStandard}). They are compared to the confidence intervals obtained with the CMC method on \(10^7\) trajectories.}
    \label{fig:compareIC}
\end{center}
\end{figure}

We carry out three series of numerical simulations on the spent fuel pool system.
\begin{enumerate}
    \item We first compare the performance of each version of our AIS method to a CMC method on a standard case with jump rates described in \cref{tab:standardCase} in appendix, results described in \cref{tab:resultsStandard} and a probability of system failure about \(10^{-5}\).
    \item We then check the stability of the best version of our method which seems to be based on \(\Ua{(MPS)}\). We represent in \cref{fig:compareIC} 50 confidence intervals at 95\% level obtained with the AIS MPS-IF method with a sample size of \(10^3\) trajectories still on the standard case (\cref{tab:standardCase}) and we compare them to the confidence interval obtained with the CMC method and a sample size of \(10^7\). 
    \item Since the method is stable, we can trust the confidence intervals produced and confront it with even rarer events for which it cannot be compared to a CMC method. Therefore we test the AIS MPS-IF method on an extreme case with jump rates described in \cref{tab:extremeCase} in appendix, results described in \cref{tab:resultsExtreme} and a probability of system failure about \(10^{-7}\).
\end{enumerate}

\begin{table}

\begin{center}
\renewcommand{\arraystretch}{1.3}
{\footnotesize
\begin{tabular}{|c | c | c | c | c | }
\hline
\textbf{Method} & \(N_L\) &  \textcolor{changes}{\(\widehat{P}_{N_L}\)} & \textcolor{changes}{\(\widehat{\sigma}_{N_L}\)} & 95\% confidence interval  \\
\hline \hline 
 & \(10^5\) & \(2 \times 10^{-5}\) &  $4.47 \times 10^{-3}$ & \(\left[ 0 \, ; \, 4.77\times 10^{-5} \right]\) \\
CMC & \(10^6\) & \(1.3 \times 10^{-5}\) &  $3.61 \times 10^{-3}$ & \(\left[ 5.93 \times 10^{-6} \, ; \, 2.01\times 10^{-5} \right]\)\\
 & \(10^7\) & \(1.77 \times 10^{-5}\) &  $4.21 \times 10^{-3}$ & \(\left[ 1.51 \times 10^{-5} \, ; \, 2.03\times 10^{-5} \right]\) \\
\hline 
\multirow{2}{*}{AIS with \(\Ua{(BC)}\)} & \(10^3\) &  \(2.16 \times 10^{-5}\) & $2.35 \times 10^{-4}$ & \(\left[ 7.05\times 10^{-6} \, ; \, 3.63\times 10^{-5} \right]\) \\
 & \(10^4\) & \(1.79 \times 10^{-5}\)
 & $3.01 \times 10^{-4}$ & \(\left[ 1.37\times 10^{-5} \, ; \, 2.22\times 10^{-5} \right]\) \\
 \hline 
\multirow{2}{*}{AIS with \(\Ua{(MPS)}\)} & \(10^3\) & \(2.19\times 10^{-5}\) & $6.59 \times 10^{-5}$ & \(\left[1.78 \times 10^{-5} \, ; \, 2.60 \times 10^{-5}\right]\) \\
 & \(10^4\) & \(1.99\times 10^{-5}\) & $2.01 \times 10^{-5}$ & \(\left[1.96 \times 10^{-5} \, ; \, 2.03 \times 10^{-5}\right]\)\\
 \hline
\multirow{2}{*}{AIS with \(\Ua{(MCS)}\)} & \(10^3\) & \(1.05\times 10^{-5}\) & $1.27 \times 10^{-4}$ & \(\left[ 2.65\times 10^{-6} \, ; \, 1.83 \times 10^{-5} \right]\)\\
 & \(10^4\) & \(1.50\times 10^{-5}\) & $2.04 \times 10^{-4}$ & \(\left[1.10 \times 10^{-5} \, ; \, 1.90 \times 10^{-5}\right]\)\\
 \hline 

\end{tabular}

}
\caption{\footnotesize\label{tab:resultsStandard}\textbf{Results on the standard SFP case} (with jump rates from \cref{tab:standardCase}). The 3 AIS methods were initialized according to the same way described in \cref{sec:Implement}: with the smallest one-dimensional \(\bm{\theta}\) such that the probability that at least one component failure occurs before the end of the simulation is larger than \(1/3\). At each iteration, we generate trajectories until we have \(n_{\text{CE}} = 10\) failures before updating \(\bm{\theta}\) for \(N = 10^3\) and \(n_{\text{CE}} = 50\) for \(N = 10^4\). We stop when the total budget \(N\) is reached. The effective dimension of the vector \(\bm{\theta}\) is reduced to 8. For the CMC method, we simply generate \(N\) trajectories and we count the proportion of faulty trajectories. The estimated probability \({\widehat{P}_{N_L}}\) is given by \cref{eq:EstimateP} and the estimated asymptotic standard deviation \({\widehat{\sigma}_{N_L}}\) is  given by \cref{eq:Var}.}
\end{center}
\end{table} 

The first observation on \cref{tab:resultsStandard} is that even the BC-IF method, which does not distinguish the role of each component, manages to drastically reduce the variance of the estimator compared to the CMC method (almost by a factor of 1000). Surprisingly, the performance of the MCS-IF method is closer to the BC-IF method than to the MPS-IF method. The latter is extremely efficient with a variance reduction of \(10^4\). We may explain this by two reasons. The first one is that, as we have seen, the MPS-IF method is more adapted to parallel systems than the MCS-IF method, yet the structure of reliable industrial systems relies on the redundancy of components and thus on parallelism. The second one is that since we are dealing with a dynamic system in continuous time and not in discrete time, it seems more appropriate to decide how fast to go through the stages until \(\mathbb{M}_{D}\) is reached, as the MPS-IF method allows, rather than to decide which stages are to be gone through in priority, as the MCS-IF method allows. 

Note that the coefficient of variation in \cref{tab:resultsStandard}, which can be used as an indicator of the performance of an estimator, does not necessarily decrease with the sample size. This is due to the fact that in the first iterations of the method, the parameter \(\bm{\theta}\) is not yet well chosen and that a poor importance distribution in high dimension tends to produce too small likelihood ratios. The coefficient of variation is underestimated at this time. 

\Cref{fig:compareIC} confirms the performance of the AIS MPS-IF method. The majority of the confidence intervals produced by the AIS method with a sample size of \(10^3\) are shorter than the confidence interval produced by the CMC method with a sample size of \(10^7\). Only 1 interval out of 50 is significantly larger than the interval produced by CMC, but it is  relevant since it gives a probability of failure between \(1\times 10^{-5}\) and \(3.5\times 10^{-5}\). We deduce that the AIS MPS-IF method is robust and that we can therefore have confidence in its estimates. 

\begin{table}

\begin{center}
\renewcommand{\arraystretch}{1.3}
{\footnotesize
\begin{tabular}{|c | c | c | c | c | }
\hline
\textbf{Method} & \(N_L\) &  \textcolor{changes}{\(\widehat{P}_{N_L}\)} & \textcolor{changes}{\(\widehat{\sigma}_{N_L}\)} & 95\% confidence interval  \\
\hline \hline 
\multirow{2}{*}{AIS with \(\Ua{(MPS)}\)} & \(10^3\) & \(3.31 \times 10^{-7}\) & $1.11 \times 10^{-6}$ & \(\left[ 2.63\times 10^{-7} \, ; \, 4.00 \times 10^{-7} \right]\)\\
 & \(10^4\) & \(3.83\times 10^{-7}\) & $1.26 \times 10^{-6}$ & \(\left[3.58 \times 10^{-7} \, ; \, 4.08 \times 10^{-7}\right]\)\\
 \hline

\end{tabular}
}
\caption{\footnotesize\label{tab:resultsExtreme}\textbf{Results with the MPS-IF approximation on the extreme SFP case} (with jump rates from \cref{tab:extremeCase}). 
Same method as in Table \ref{tab:resultsStandard} except that
the initialization follows the method described in \cref{sec:Implement} with the smallest one-dimensional \(\bm{\theta}\) such that the probability that at least one component failure occurs before the end of the simulation is larger than $0.9$. 
}
\end{center}
\end{table} 

Finally, we see on the \cref{tab:resultsExtreme} that the AIS MPS-IF method still offers excellent performances for a 100 times rarer event. A reliable estimate of the probability that is of order \(10^{-7}\) can be obtained  with a sample size smaller than \(10^4\).

\section{Conclusion}
\label{sec:End}

This work contains a comprehensive methodology for assessing the reliability of hybrid dynamic industrial systems, as well as a demonstration of its efficiency.
\begin{enumerate}
    \item We have presented the mathematical modeling of the system under the form of a piecewise deterministic Markov process (PDMP).
    \item We have emphasized the role played by the committor function of the system in the optimality conditions of an importance sampling method for estimating its failure probability. 
    \item We have proposed three different families of parametric approximations of the committor function. The forms of these families are based on the decomposition of the system structure function into minimal path sets (MPS) and minimal cut sets (MCS). 
    \item We have proposed
    an adaptive importance sampling (AIS) algorithm based on a cross-entropy procedure and a recycling scheme of past samples. The convergence and asymptotic normality of the estimator have been demonstrated. They make it possible to construct asymptotic confidence intervals of the failure probability. 
    \item Finally, the different versions of our method have been tested and compared on different test cases.
    
\end{enumerate}

It is found that each version of our AIS method is considerably more efficient than a CMC method in all cases. If we compare the different AIS versions between them, it appears that the best performances are obtained when we approximate the committor function with an increasing function in the number of MPS with a broken component. The variance of the estimator produced is more than 10,000 times smaller than that of a CMC method on the examples. It allows to estimate with accuracy a probability of failure of order \(10^{-7}\) with a sample size smaller than \(10^4\).\\

\color{changes}
\paragraph{Multi-level CE and improved CE} Sometimes, it is challenging to determine an initial instrumental distribution beforehand that enables the realization of the event $\left\{\mathcal{Z}\in\mathcal{D}\right\}$. When the event takes the form $\left\{\varphi(\mathcal{Z}) > \overline{\gamma}\right\}$, a common technique is to adaptively set intermediate thresholds $\gamma_1 < \gamma_2 < \dots < \overline{\gamma}$ and replace the indicator $\mathds{1}_{\varphi(\mathcal{Z})>\overline{\gamma}}$ with $\mathds{1}_{\varphi(\mathcal{Z})>\gamma_\ell}$ at each step $\ell$, as explained in Algorithm 1.1 of \cite{deboerTutorialCrossEntropyMethod2005}. A further refinement of this technique is the iCE (improved cross entropy method) presented in \cite{uribeCrossentropybasedImportanceSampling2020}, where the indicator function is replaced by a continuous approximation of the form $g(\mathcal{Z} ; s) = F_{\mathcal{N}(0,1)}\left(\frac{\varphi(\mathcal{Z})-\overline{\gamma}}{s}\right)$ with $F_{\mathcal{N}(0,1)}$ the cumulative distribution function of the $\mathcal{N}(0,1)$ distribution (such that we have $g(\mathcal{Z} ; s) \underset{s\rightarrow 0}{\longrightarrow} \mathds{1}_{\varphi(\mathcal{Z})>\overline{\gamma}}$).

In our case, as we have seen, even though the event $\left\{X_t \in \mathbb{X}_D\right\}$ can generally be expressed as a threshold exceedance, the intermediate steps to be crossed are primarily determined by the modes of $\mathbb{M}$ that are not ordered. The importance function can serve both to parameterize the importance distribution and to define the intermediate thresholds by ordering the modes. The major drawback of MCS-IF here is that it classifies the modes in a different order depending on the value of the vector $\bm{\theta}$ (unlike BC-IF and MPS-IF).\\

\color{black}
\paragraph{Extension to other applications with reverse importance sampling trick} Our method can also serve other purposes. Recall that with the reverse importance trick, one can always estimate what the probability of failure would have been under another distribution \(\widetilde{\pi}\):
\begin{equation}
    \mathbb{E}_{\widetilde{\pi}}\left[\mathds{1}_{\mathcal{Z}\in\mathcal{D}}\right] = \mathbb{E}_{\qtheta}\left[\mathds{1}_{\mathcal{Z}\in\mathcal{D}} \dfrac{\widetilde{\pi}(\mathcal{Z})}{\qtheta(\mathcal{Z})}\right]  .
\end{equation}
For example, a reliability sensitivity analysis can be carried out \color{changes} to measure the influence of variations of the jump intensity and the jump kernel (or hyperparameters of the jump intensity and the jump kernel such as failure rates of components of an industrial system) on its failure probability. Any classical sensitivity index \cite{razavi2021future} can be constructed from an input/output data set \(\left((\lambda,K)^{(i)},\widehat{P}_{\pi_{(\lambda,K)^{(i)}}}\right)_{i=1,\dots,n}\) with: 
\begin{equation}\widehat{P}_{\pi_{(\lambda,K)^{(i)}}} =
\dfrac{1}{N_{L}} \sum_{\ell=1}^L \sum_{k=1}^{n_{\ell}} \mathds{1}_{\mathcal{Z}_k^{(\ell)}\in\mathcal{D}} \frac{\pi_{(\lambda,K)^{(i)}}\big(\mathcal{Z}_k^{(\ell)}\big)}{q_{\bm{\theta}^{(\ell)}}\big(\mathcal{Z}_k^{(\ell)}\big) } ,
\end{equation}
for $i=1,\ldots,n$. 
\color{black}
Thus the trajectories already simulated can be recycled to estimate new quantities. \\
%


\paragraph{Application to other rare event methods} \color{changes} As mentioned earlier, approximating the committor function of the process enables the efficient implementation of variance reduction methods other than importance sampling. Importance splitting is a family of methods used to estimate the probability of a rare event by decomposing it into a nested intersection of less rare events. The principle is to generate a set of trajectories of the process, this time following its original distribution $\pi_0$, but duplicating the most promising trajectories along the way and discarding the others. It is up to the user to choose an importance function that determines whether a trajectory is promising or not, and this choice primarily determines the method's performance.
Such methods have already been applied to PDMPs in the literature. For example, adaptive multilevel splitting (AMS) \cite{cerou07,brehier16} was applied to particle transport in \cite{louvin2017adaptive}, and the interacting particle systems (IPS) method \cite{delmo05} was applied to industrial systems similar to ours in  \cite{chraibiOptimalPotentialFunctions}. 
The optimal importance function to use in AMS is the committor function $U_{\text{opt}}$. In the case of the IPS algorithm, the optimal importance function (more exactly, the potential function used to select the promising particles) can also be expressed in terms of the committor function although the relationship is more complex. 
The families of importance functions we have proposed in this paper could therefore be used to efficiently implement splitting algorithms. The latter do not generally compete with a well-implemented importance sampling method, however their performance degrades little when the importance function is not well chosen. They generally require less a priori knowledge about the system.

\color{black}
\bibliographystyle{siamplain}

\bibliography{references} 
\appendix
\section{Appendix}

\subsection{Proof of \cref{th:Asymp}}
\label{subsec:Proof}
The results come directly from showing that we verify the hypotheses of Theorems 2 and 3 from \cite{delyonAsymptoticOptimalityAdaptive2018}. We know by \cref{hyp:PDMP,hyp:U_theta}  that \(\Theta\) is compact and that we have \(\mathbb{E}_{\pi_{0}} \left[-\mathds{1}_{\mathcal{Z}\in\mathcal{D}} \log \qtheta(\mathcal{Z})\right] > \mathbb{E}_{\pi_{0}}\left[-\mathds{1}_{\mathcal{Z}\in\mathcal{D}} \log q_{\thetaopt}(\mathcal{Z})\right]\) if \(\bm{\theta} \neq \thetaopt\).
Moreover, for any \(z\in E\) the continuity of the application \(\bm{\theta} \mapsto \Utheta(z)\) implies the continuity of \(\bm{\theta} \mapsto \mathds{1}_{\mathcal{Z}\in\mathcal{D}}\, \pi_{0}(\mathcal{Z}) \log \qtheta(\mathcal{Z})\) for any \(\mathcal{Z}\in\mathcal{E}\).
To obtain the convergence of the sequence \(\left(\bm{\theta}^{(L)}\right)_{L>0}\), it remains to show that~:
 \begin{align}
 \label{eq:Cond1LGN}
    &\mathbb{E}_{\pi_{0}}\left[\mathds{1}_{\mathcal{Z}\in\mathcal{D}} \sup_{\tT}\left\{-\log \qtheta(\mathcal{Z}) \right\} \right] < +\infty ,\\
\label{eq:Cond2LGN}
    &\sup_{\tT}\left\{\mathbb{E}_{\pi_{0}}\left[\mathds{1}_{\mathcal{Z}\in\mathcal{D}} \dfrac{\pi_{0}(\mathcal{Z})}{\qtheta(\mathcal{Z})} \sup_{\widetilde{\bm{\theta}}\in\Theta} \left\{-\log \pi_{\widetilde{\bm{\theta}}}(\mathcal{Z}) \right\}^2 \right]\right\} < +\infty , 
\end{align}
and to get the asymptotic normality of the estimator \(\widehat{P}_{N_L}\), we have to prove that there exists \(\eta>0\) such that:
\begin{equation}
\label{eq:CondTCL}
    \sup_{\tT} \left\{\mathbb{E}_{\pi_{0}}\left[\mathds{1}_{\mathcal{Z}\in\mathcal{D}} \left(\dfrac{\pi_{0}(\mathcal{Z})}{\qtheta(\mathcal{Z})}\right)^{1+\eta} \right]\right\} < +\infty .
\end{equation}

From the definitions \cref{eq:thetaIntKer} of \(\lambdatheta\) and \(\Ktheta\), and from \cref{hyp:PDMP,hyp:U_theta}, we obtain that for any \(z\in \text{supp}\left(\lambdazero\right)\):
$
    \lambda_{\min} {U_{\min}}/{U_{\max}} \leq \lambdatheta(z)  \leq \lambda_{\max} {U_{\max}}/{U_{\min}}
$,
and for any \(z^-\in E\) and any \(z \in \text{supp} \left(\Kzero\left(z^-,\cdot \right)\right)\):
$
    K_{\min} {U_{\min}}/{U_{\max}} \leq K(z^-,z)  \leq K_{\max} {U_{\max}}/{U_{\min}}
$.
Then from the definition \cref{eq:Density} of the density of a PDMP trajectory \(\mathcal{Z} \in \mathcal{E}\) with \(\nZ\) jumps,  there exist \(c_{\min},c_{\max}>0\) such that:
\begin{align}
\label{eq:SupDens}
\sup_{\bm{\theta}\in\Theta} \qtheta(\mathcal{Z}) &\leq \prod_{k=0}^{\nZ} \left(\lambda_{\max} \dfrac{U_{\max}}{U_{\min}}\right)^{\mathds{1}_{t_k<t^{\partial}_{z_k}}} \prod_{k=0}^{\nZ - 1} K_{\max}\dfrac{U_{\max}}{U_{\min}} \leq \left(c_{\max}\right)^{\nZ} , \\
\label{eq:InfDens}
\inf_{\bm{\theta}\in\Theta} \qtheta(\mathcal{Z}) &\geq \prod_{k=0}^{\nZ} \left(\lambda_{\min} \dfrac{U_{\min}}{U_{\max}}\right)^{\mathds{1}_{t_k<t^{\partial}_{z_k}}} \exp\left[ -\lambda_{\max} \dfrac{U_{\max}}{U_{\min}}  t_k \right] \prod_{k=0}^{\nZ - 1} K_{\min} \dfrac{U_{\min}}{U_{\max}} \geq \left(c_{\min}\right)^{\nZ} .
\end{align}

Using \cref{eq:SupDens,eq:InfDens} we see that conditions \cref{eq:Cond1LGN,eq:Cond2LGN,eq:CondTCL} are dominated by the following: for any constant \(c>0\), 
$
    \mathbb{E}_{\pi_{0}}\left[c^{\nZ} \right] < +\infty
$.
We have \(\nZ = n^{(\lambda)}_{\mathcal{Z}} + n^{(\partial)}_{\mathcal{Z}}\) with \(n^{(\lambda)}_{\mathcal{Z}}\) the number of spontaneous jumps with jump rate \(\lambda\) and \(n^{(\partial)}_{\mathcal{Z}}\) the number of jumps at boundaries. At most, the process reaches the state space boundary "almost immediately" after each spontaneous jump, and a time \(t_{\varepsilon}\) after reaching another boundary. So \(n^{(\partial)}_{\mathcal{Z}} \leq n^{(\lambda)}_{\mathcal{Z}} + t_{\max}/t_{\varepsilon}\), and thus \(\nZ \leq 2 n^{(\lambda)}_{\mathcal{Z}} + t_{\max}/t_{\varepsilon}\). We just need to prove that \( \mathbb{E}_{\pi_{0}}\left[c^{n^{(\lambda)}_{\mathcal{Z}}} \right] < +\infty\) for any constant \(c\). 
 
We define \(\widetilde{\mathcal{Z}}\) as a jump process analogous to a PDMP but with some jumps rejected and not taking place. It is characterized by its flow \(\Phi\), its constant jump intensity \(\lambda_{\max}\) and its jump kernel \(\widetilde{\mathcal{K}}\) defined as follows:
\begin{equation}
       \widetilde{\mathcal{K}}\left(z^-,dz\right) =  \left\{
    \begin{array}{ll}
      \mathcal{K}_{0}\left(z^-,dz\right) \dfrac{\lambdazero(z^-)}{\lambda_{\max}} + \left(1- \dfrac{\lambdazero(z^-)}{\lambda_{\max}}\right) \delta_{z^-}(dz)  & \mbox{if } z^- \notin \partial E, \\
        \mathcal{K}_{0}\left(z^-,dz\right)   & \mbox{otherwise.} 
    \end{array}
\right. 
\end{equation}

Thus a part of the spontaneous jumps are "rejected" because the process remains on the same state at each jump with probability \(\left(1- \frac{\lambdazero(z^-)}{\lambda_{\max}}\right)\). Following theorem 5.5 from \cite{davisPiecewiseDeterministicMarkovProcesses1984}, we notice that the generator of this process is the same as that of the PDMP \(\mathcal{Z}\). Indeed by denoting \(\mathcal{Q}_{0}\) the generator of the PDMP \(\mathcal{Z}\) and \(\widetilde{\mathcal{Q}}\) the generator of the process \(\widetilde{\mathcal{Z}}\), for any state \(z^-\in E\) and \(f\) a function of the domain of the generator \(\widetilde{\mathcal{Q}}\) (see detail in \cite{davisPiecewiseDeterministicMarkovProcesses1984}):
\begin{align*}
    &\widetilde{\mathcal{Q}} \,f(z^-) = \langle \nabla f(z^-),\bm{g}(z^-)\rangle + \lambda_{\max}(z^-)\int_E \left[f(z) - f(z^-) \right] \widetilde{\mathcal{K}}(z^-,dz)  \\
    &= \langle \nabla f(z^-),\bm{g}(z^-)\rangle + \lambdazero(z^-)\int_E \left[f(z) - f(z^-) \right] \mathcal{K}_{0}(z^-,dz) + \left(1- \frac{\lambdazero(z^-)}{\lambda_{\max}}\right) \left[f(z^-) - f(z^-) \right] \\
    &= \mathcal{Q}_{0} \,f(z^-).
\end{align*}

Since the generator characterizes the distribution of the process, the trajectories of the PDMP \(\mathcal{Z}\) and of the jump process \(\widetilde{\mathcal{Z}}\) are identically distributed. In particular, their number of jumps \(\nZ\) and \(n_{\widetilde{\mathcal{Z}}}\) also follow the same law (as well as \(\nZ^{(\lambda)}\) and \(n_{\widetilde{\mathcal{Z}}}^{(\lambda)}\)). If we note \(\widetilde{n}_{\widetilde{\mathcal{Z}}}^{(\lambda)}\) the number of proposed jumps with jump intensity \(\lambda_{\max}\) including the rejected ones, it is straightforward to see that \(\widetilde{n}_{\widetilde{\mathcal{Z}}}^{(\lambda)}\) follows a Poisson distribution with intensity \(\lambda_{\max}\) and that \(\widetilde{n}_{\widetilde{\mathcal{Z}}}^{(\lambda)}\geq n_{\widetilde{\mathcal{Z}}}^{(\lambda)}\). Finally for any \(c\geq 1\),
\begin{equation}
\mathbb{E}_{\pi_{0}}\left[c^{n^{(\lambda)}_{\mathcal{Z}}} \right] = \mathbb{E}\left[c^{n^{(\lambda)}_{\widetilde{\mathcal{Z}}}} \right] \leq \mathbb{E}\left[c^{\widetilde{n}^{(\lambda)}_{\widetilde{\mathcal{Z}}}} \right] = \mathbb{E}\left[c^{\mathcal{P}(\lambda_{\max})} \right] = e^{\lambda_{\max}(c-1)} < +\infty  .
\end{equation}
This completes the proof of the theorem.
 \qed

\subsection{Gradient of the log-likelihood for instrumental distributions}
\label{subsec:Grad}
At each iteration of the cross-entropy procedure, the minimization program \cref{eq:MinCE} must be solved. The only quantity depending on \(\bm{\theta}\) in the objective function is : \(\mathcal{Z} \mapsto \log \qtheta(\mathcal{Z})\). Let us recall that the probability density function of any trajectory \(\mathcal{Z}\in \mathcal{E}\) is given by \cref{eq:Density}. For all states \(z^-,z \in E\), we note: \(r_{\bm{\theta}}(z^-,z) = \Utheta^-(z^-)/\Utheta(z)\). For \(i\in\{1,\dots,d_{\bm{\theta}}\}\), the derivative of \(\log \qtheta(\mathcal{Z})\) in \(\theta_i\) is given by:
\begin{multline*}
    \partial_{\theta_i} \log \qtheta(\mathcal{Z}) = \sum_{k=0}^{\nZ} \left[\mathds{1}_{t_{z_k} < t_{z_k}^{\partial}} \dfrac{\partial_{\theta_i} r_{\bm{\theta}}\left(\Phi_{z_k}(t_k),\Phi_{z_k}(t_k)\right)}{r_{\bm{\theta}}\left(\Phi_{z_k}(t_k),\Phi_{z_k}(t_k)\right)} \right. \\
    \left. - \int_0^{t_k} \lambdazero\left(\Phi_{z_k}(u)\right) \partial_{\theta_i} r_{\bm{\theta}}\left(\Phi_{z_k}(u),\Phi_{z_k}(u)\right) du \right]  
    - \sum_{k=0}^{\nZ-1} \dfrac{\partial_{\theta_i} r_{\bm{\theta}}\left(\Phi_{z_k}(t_k),z_{k+1}\right)}{r_{\bm{\theta}}\left(\Phi_{z_k}(t_k),z_{k+1}\right)}  ,
\end{multline*}
with $ \displaystyle \partial_{\theta_i} r_{\bm{\theta}}(z^-,z) = \dfrac{1}{\Utheta(z)^2} \int_E \left[\Utheta(z) \partial_{\theta_i} \Utheta(z^+) -  \Utheta(z^+) \partial_{\theta_i} \Utheta(z)\right] \Kzero(z^-,z^+)\, d\nu_{z^-}(z^+)$.

\subsection{Jump parameters for the series/parallel systems and the spent fuel pool system}

The marginal jump rates of each system component are presented according to its status and according to the value of the position. The jump intensity of the process in a given state is the sum of the marginal jump rates in that state.

\vspace{0.4in}

\begin{table}[h!]

\begin{center}
\renewcommand{\arraystretch}{1.1}
{\footnotesize
\begin{tabular}{|c | c | c |}
\cline{2-3}  
\multicolumn{1}{c|}{} & \textbf{Series system} & \textbf{Parallel system} \\  
\hline 
\textbf{Component} & \multicolumn{2}{c|}{Marginal jump intensity \(\lambda^{(j)}_{0}\) for  \(j=1,\dots,d_{\bm{c}}\)} \\
\hline \hline
\(c_{1}\) & \(1\cdot 10^{-9}\,\mathds{1}_{m^{(1)} = 0} + 1\cdot 10^{-6}\,\mathds{1}_{m^{(1)} = 1}\) &\(6\cdot 10^{-5}\,\mathds{1}_{m^{(1)} = 0} + 1\cdot 10^{-4}\,\mathds{1}_{m^{(1)} = 1}\) \\ 

\(c_{2}\) & \(5\cdot 10^{-9}\,\mathds{1}_{m^{(2)} = 0} + 5\cdot 10^{-6}\,\mathds{1}_{m^{(2)} = 1}\)  &\(2\cdot 10^{-4}\,\mathds{1}_{m^{(2)} = 0} + 5\cdot 10^{-4}\,\mathds{1}_{m^{(2)} = 1}\)  \\

\(c_{3}\) & \(5\cdot 10^{-9}\,\mathds{1}_{m^{(3)} = 0} + 1\cdot 10^{-6}\,\mathds{1}_{m^{(3)} = 1}\) &\(2\cdot 10^{-4}\,\mathds{1}_{m^{(3)} = 0} + 1\cdot 10^{-3}\,\mathds{1}_{m^{(3)} = 1}\)     \\ 

\(c_{4}\) & \(1\cdot 10^{-9}\,\mathds{1}_{m^{(4)} = 0} + 5\cdot 10^{-6}\,\mathds{1}_{m^{(4)} = 1}\) &\(6\cdot 10^{-5}\,\mathds{1}_{m^{(4)} = 0} + 5\cdot 10^{-4}\,\mathds{1}_{m^{(4)} = 1}\)    \\ 

\(c_{5}\) & \(8\cdot 10^{-9}\,\mathds{1}_{m^{(5)} = 0} + 6\cdot 10^{-6}\,\mathds{1}_{m^{(5)} = 1}\) &\(5\cdot 10^{-4}\,\mathds{1}_{m^{(5)} = 0} + 8\cdot 10^{-4}\,\mathds{1}_{m^{(5)} = 1}\)  \\

\hline
\end{tabular}
}
\caption{\footnotesize\label{tab:seriesParallelParameters} Marginal jump intensity of each component for the \textbf{series} and \textbf{parallel} systems.}
\end{center}
\end{table}


\begin{table}[H]

\begin{center}
\renewcommand{\arraystretch}{1.1}
{\footnotesize
\begin{tabular}{|c | c | c | c | }
\hline
\textbf{Component} & \multicolumn{3}{c|}{Marginal jump intensity \(\lambda^{(j)}_{0}\) for  \(j=1,\dots,d_{\bm{c}}\)} \\
\hline
\(c_i\) & when \(m^{(i)} = -1 \) & when \(m^{(i)} = 0 \) & when \(m^{(i)} = 1 \) \\
\hline \hline
\(c_1 = \texttt{G}_{0}\) & \(4\cdot 10^{-2}\) & \(4\cdot 10^{-6}\) & \(6\cdot 10^{-6}\) \\ 

\(c_{i+1} =\texttt{G}_{i}\), $i=1,2,3$ & \(8\cdot 10^{-2}\) & \(2\cdot 10^{-6}\) & \(30\cdot 10^{-6}\) \\




\(c_5 = \texttt{S}_{1}\) & \(1\cdot 10^{-2}\) &  \(4\cdot 10^{-6}\) & \(20\cdot 10^{-6}\)  \\

\(c_6 = \texttt{S}_{2}\) & \(3\cdot 10^{-2}\) & \(1\cdot 10^{-6}\) & \(5\cdot 10^{-6}\) \\

\(c_{6+i} = \texttt{L}_{i,1}\), $i=1,2,3$ & \((6 - 0.03 X^{(1)}_t)\cdot 10^{-2}\) & \((1 + 0.05 X^{(1)}_t)\cdot 10^{-6}\) & \((3 + 0.1 X^{(1)}_t)\cdot 10^{-6}\) \\




\(c_{9+i} = \texttt{L}_{i,2}\), $i=1,2,3$ & \((6 - 0.03 X^{(1)}_t)\cdot 10^{-2}\) & \((1 + 0.05 X^{(1)}_t)\cdot 10^{-6}\) & \((3 + 0.1 X^{(1)}_t)\cdot 10^{-6}\) \\




\(c_{12+i} = \texttt{L}_{i,3}\), $i=1,2,3$ & \((6 - 0.03 X^{(1)}_t)\cdot 10^{-2}\) & \((1 + 0.05 X^{(1)}_t)\cdot 10^{-6}\) & \((3 + 0.1 X^{(1)}_t)\cdot 10^{-6}\) \\



\hline
\end{tabular}
}
\caption{\footnotesize\label{tab:standardCase} Marginal jump intensity of each component  for the \textbf{standard SFP case}.}
\end{center}
\end{table}


\begin{table}[H]

\begin{center}
\renewcommand{\arraystretch}{1.1}
{\footnotesize
\begin{tabular}{|c | c | c | c | }
\hline
\textbf{Component} & \multicolumn{3}{c|}{Marginal jump intensity \(\lambda^{(j)}_{0}\) for  \(j=1,\dots,d_{\bm{c}}\)} \\
\hline
\(c_i\) & when \(m^{(i)} = -1 \) & when \(m^{(i)} = 0 \) & when \(m^{(i)} = 1 \) \\
\hline \hline
\(c_1 = \texttt{G}_{0}\) & \(4\cdot 10^{-2}\) & \(4\cdot 10^{-6}\) & \(6\cdot 10^{-6}\) \\ 

\(c_{i+1} =\texttt{G}_{i}\), $i=1,2,3$ & \(10\cdot 10^{-2}\) & \(15\cdot 10^{-6}\) & \(30\cdot 10^{-6}\) \\




\(c_5 = \texttt{S}_{1}\) & \(1\cdot 10^{-2}\) &  \(4\cdot 10^{-6}\) & \(20\cdot 10^{-6}\)  \\

\(c_6 = \texttt{S}_{2}\) & \(3\cdot 10^{-2}\) & \(1\cdot 10^{-6}\) & \(5\cdot 10^{-6}\) \\

\(c_{6+i} = \texttt{L}_{i,1}\), $i=1,2,3$ & \((12 - 0.04 X^{(1)}_t)\cdot 10^{-2}\) & \((1 + 0.1 X^{(1)}_t)\cdot 10^{-6}\) & \((3 + 0.1 X^{(1)}_t)\cdot 10^{-6}\) \\




\(c_{9+i} = \texttt{L}_{i,2}\), $i=1,2,3$ & \((12 - 0.04 X^{(1)}_t)\cdot 10^{-2}\) & \((1 + 0.1 X^{(1)}_t)\cdot 10^{-6}\) & \((3 + 0.1 X^{(1)}_t)\cdot 10^{-6}\) \\




\(c_{12+i} = \texttt{L}_{i,3}\), $i=1,2,3$ &\((15 - 0.05 X^{(1)}_t)\cdot 10^{-2}\) & \( (1 + 0.08 X^{(1)}_t)\cdot 10^{-6}\) & \( (3 + 0.08 X^{(1)}_t)\cdot 10^{-6}\) \\



\hline
\end{tabular}
}
\caption{\footnotesize\label{tab:extremeCase}Marginal jump intensity of each component for the \textbf{extreme SFP case}.}
\end{center}
\end{table}

\end{document}